# Outage Behavior of Discrete Memoryless Channels Under Channel Estimation Errors

Pablo Piantanida, *Member, IEEE,* Gerald Matz, *Member, IEEE,*

and Pierre Duhamel, *Fellow, IEEE*

## Abstract

Communication systems are usually designed by assuming perfect channel state information (CSI). However, in many practical scenarios, only a noisy estimate of the channel is available, which may strongly differ from the true channel. This imperfect CSI scenario is addressed by introducing the notion of *estimation-induced outage (EIO) capacity*. We derive a single-letter characterization of the maximal EIO rate and prove an associated coding theorem and its strong converse for discrete memoryless channels (DMCs). The transmitter and the receiver rely on the channel estimate and the statistics of the estimate to construct codes that guarantee reliable communication with a certain outage probability. This ensures that in the non-outage case the transmission meets the target rate with small error probability, irrespective of the quality of the channel estimate. Applications of the EIO capacity to a single-antenna (non-ergodic) Ricean fading channel are considered. The EIO capacity for this case is compared to the EIO rates of a communication system in which the receiver decodes by using a mismatched ML

This research is supported in part by the FP7 Network of Excellence in Wireless COMmunications NEWCOM.

P. Piantanida is with the Department of Telecommunications, SUPELEC, 91192 Gif-sur-Yvette, France, Email: pablo.piantanida@supelec.fr. G. Matz is with the Institute of Communications and Radio-Frequency Engineering, Vienna University of Technology, A-1040 Wien, Austria, Email: g.matz@ieee.org. P. Duhamel is with the Laboratoire des Signaux et Systemes, CNRS-SUPELEC, 91192 Gif-sur-Yvette, France, Email: pierre.duhamel@lss.supelec.fr.

decoder. The effects of rate-limited feedback to provide the transmitter with quantized CSI are also investigated.

## I. INTRODUCTION

Channel uncertainty caused by time variations/fading, interference, or channel estimation errors, can severely impair the performance of wireless systems. Even if the channel is quasi-static and the interference is small, uncertainty induced by imperfect channel state information (CSI) at the transmitter remains. As a consequence, studying the limits of reliable information rates in these scenarios is an important problem. Obviously, this requires some precise definition of the communication model and what one means by "reliability" in the situations of interest.

In selecting a probabilistic model for a wireless communication scenario where the channel parameters are time-varying, several factors must be considered. These include the physical and statistical nature of the channel disturbances (e.g. fading distribution, channel estimation method, practical design constraints, etc.), the information available to the transmitter and/or to the receiver and the presence of any feedback link [1]. Assume that a specific instance of a discrete memoryless state-dependent channel (DMC) with discrete input $x \in \mathscr{X}$, discrete state $s \in \mathscr{S}$ and discrete output $y \in \mathscr{Y}$ is characterized by a set of conditional probability distributions (PDs) $\mathcal{W}_{\Theta} = \{W_{\theta} : \mathscr{X} \times \mathscr{S} \longmapsto \mathscr{Y}\}_{\theta \in \Theta}$, parameterized by the vector $\theta \in \Theta$, where $\Theta$ is a set of parameters (not necessarily finite). The transition PD of the $n$-memoryless extension with inputs $\mathbf{x} = (x_1, \ldots, x_n)$, channel states $(\theta, \mathbf{s}) = (\theta, s_1, \ldots, s_n)$ and outputs $\mathbf{y} = (y_1, \ldots, y_n)$ is given by

$$W_\theta^n(\mathbf{y}|\mathbf{x}, \mathbf{s}) = \prod_{i=1}^{n} W_\theta(y_i|x_i, s_i), \quad (1)$$

where $\theta$ is assumed to be fixed during the communication with PD $\mu_{\boldsymbol{\theta}}$, but $s_i$ varies from letter to letter drawn independent identically distributed (i.i.d.) from $\mu_{\boldsymbol{S}|\boldsymbol{\theta}}$. This channel model is suitable for many wireless communication scenarios. The channel is said *non-ergodic* if the conditional PD does not depend on the state $s$ and *ergodic* if the conditional PD and $s$ do not depend on $\theta$. Otherwise the channel model (1) might have fixed and time-varying channel states or parameters $(\theta, \{s_i\}_{i=1}^\infty)$ during the communication.

Existing approaches dealing with channel uncertainty mainly correspond to two scenarios. The first scenario might be characterized by two facts: (i) the transmitter and the receiver are designed without the knowledge of the law governing the channel variations $\mu_{\boldsymbol{\theta S}}$ and (ii) the receiver may only dispose of $(\hat{\theta}, \{v_i\}_{i=1}^\infty)$, i.e., noisy estimates of $(\theta, \{s_i\}_{i=1}^\infty)$, while the transmitter is provided with $(\hat{\theta}, \{v_i\}_{i=1}^\infty, \theta, \{s_i\}_{i=1}^\infty)$. A reasonable approach in this case consists in using *mismatched decoders* [2]–[5] where the decoding rule is restricted to be a metric of interest, which is not necessarily matched to the channel governing the communication. A second scenario arises when the transmitter and the receiver are both assumed to be aware of the laws governing the channel variations. Let us assume first the case where the channel is *non-ergodic* ($s$ is not present), and the transmitter is aware of the true channel states but not the receiver. In this case, *universal decoders* [6] can still achieve the capacity, attaining the same performance as the maximum-likelihood (ML) decoder tuned to the true channel. Loosely speaking, an universal decoder for a parametric family of channels $\mathcal{W}_\Theta$ is a decoder independent of the specific channel in use, that nevertheless performs asymptotically as well as the ML decoder tuned to the true channel. Many families of channels admit universal decoders (see [6], [7] and [8], and references therein). Finally, we may also identify an intermediate scenario, where the transmitter only knows noisy channel estimates. Caire and Shamai [9] have studied the capacity of *ergodic* channels with imperfect CSI at the transmitter (CSIT) and/or at the receiver (CSIR), providing optimal power allocation strategies. Whereas Lapidoth and Shamai [10] studied the robustness of Gaussian codebooks and scaled nearest neighbor decoding over a flat-fading channel with respect to inaccuracies in the CSI, characterizing the performance loss that results from channel estimation errors. Additional results obtained by Lapidoth and Moser [11] show that for *non-coherent* channels (absence of CSI) the asymptotic MIMO capacity increases doubly-logarithmically with the SNR but with a reduced slope. This line of work was initiated by Marzetta and Hochwald [12], and then explored by Zheng and Tse [13], to study the non-coherent capacity under a block-fading assumption.

*A. Motivation*

The results recalled above correspond to communication scenarios where the laws governing the parameters of the channel are supposed unknown, or, whenever imperfect CSI is assumed at either the transmitter and/or the receiver, the channels are assumed *ergodic* ($\theta$ is not present). However, in many practical wireless systems operating over fading channels, the *ergodic* assumption is not necessarily satisfied since some of the channel parameters may be almost constant for a period of time, so that its randomness can not be averaged out (or removed) over time. This effect is even stronger when delay constraints are tight [14]. In such systems, a channel estimate is required for each period of time, and probably the most common method for channel estimation is the use of a training sequence. The transmitter sends a known sequence of symbols, allowing the receiver to estimate the channel state and send it back to the transmitter via a noisy feedback link. Once the channel state has been estimated, the receiver decodes the rest of the transmission. Methods for estimating the channel parameters online or refining the initial estimate have been proposed, but we do not consider them in this work. In our setting, once the initial estimate $\hat{\theta}$ has been obtained in the training phase, it is assumed fixed. This corresponds to how most actual standards work. Note that training introduces a throughput (and power) penalty since it requires frequent retransmission of the training symbols carrying no information. Hence, to reduce this undesired effect, it may be preferable to sacrifice channel estimation accuracy for small training overhead. Nevertheless, the quality-of-service (QoS) constraints must be guaranteed for each communication. This paper intends to provide insights regarding this tradeoff.

In the *non-ergodic* scenario, most communication system aim to guarantee with high probability reliable communication (small error probability) at the target rate, no matter which channel estimate arises during the communication. To this end, the system designer will use the available CSI to appropriately allocate the available resources, e.g. power for transmission, the amount of training used, etc. This scenario can be mathematically modeled as depicted in Fig. 1. A specific instance of the channel is described by a conditional PD $\{W_\theta(y|x,s)\}$ where it is assumed that neither the transmitter nor the receiver know exactly the channel states $(\theta \in \Theta, s \in \mathscr{S})$. The channel state $\theta$ is randomly drawn from $\theta \sim \mu_{\boldsymbol{\theta}}$ before the communication starts and remains unchanged throughout the transmission, while the channel state $s$ is assumed to change from letter to letter drawn i.i.d. from $\mu_{S|\boldsymbol{\theta}}$. We assume that $(s, u, v)$ is an i.i.d. sequence over $\mathscr{S} \times \mathscr{U} \times \mathscr{V}$ with joint PD $\mu_{\boldsymbol{SUV}|\boldsymbol{\theta}}$, where the transmitter is provided[1] with noisy (maybe poor) estimates $(\hat{\theta}, \{u_i\}_{i=1}^\infty)$ of $(\theta, \{s_i\}_{i=1}^\infty)$, while the receiver knows estimates $(\hat{\theta}, \{v_i\}_{i=1}^\infty)$. The encoder and the decode are both assumed to be aware of the laws governing the channel variations $(\mu_{\boldsymbol{\theta SUV}}, \mathcal{W}_\Theta)$. We remark that even in *non-ergodic* scenarios, where the channel does not have a fixed state $\theta$, the model depicted in Fig. 1 can be useful to exhibit channel uncertainty arising on the statistic controlling the time-varying states $s$. Moreover in this setting, additional information is available from the *accurate statistic*, which consists of the conditional PD $\mu_{\boldsymbol{\theta SUV}|\hat{\boldsymbol{\theta}}}$ that can be derived from the estimation method and the PDs $(\mu_{\boldsymbol{\theta SUV}}, \mathcal{W}_\Theta)$. This information can be used to measure, in terms of probabilities, how accurate the state estimates are (e.g. to compute its variance, any confidence interval, etc.). However, reliable communication cannot always be guaranteed, since extremely poor estimates (even if unlikely) are possible. We address this problem by introducing the notion of *estimation-induced outage (EIO) capacity*.

---

[1] For simplicity, we assume that $\hat{\theta}$ is available at the transmitter and at the receiver. Generalization to the case where the estimate at the transmitter would be different (due to non perfect feedback) is straightforward.

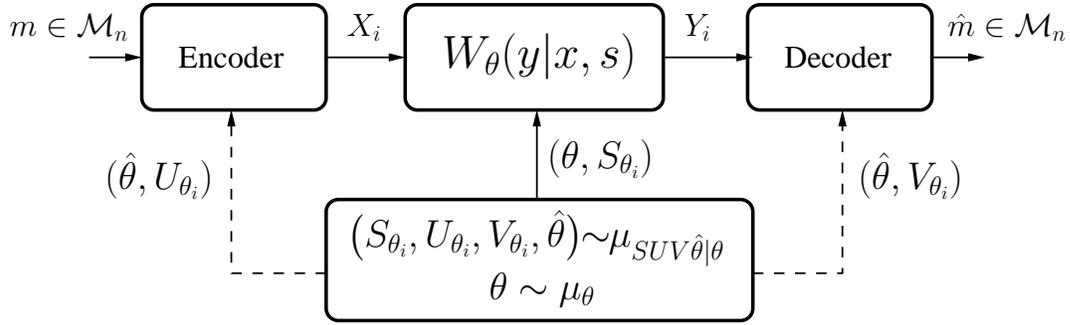

Fig. 1. Block diagram of the channel with time-varying and fixed states and imperfect CSIT and CSIR.

*B. Related Results*

We next recall further results that address closely related problems and we comment on their differences and similarities to our approach.

Médard [15] derives capacity bounds for slow fading channel with additive white Gaussian noise (AWGN) and minimum-mean square error (MMSE) channel estimation. The bounds depend on the variance of the channel estimation error regardless of the estimation method. Moreover, these results have been extended to *ergodic* fading channels in [16], [17]. More recent work by Yoo and Goldsmith [18] derives capacity lower bounds for MIMO fading channels. These capacity bounds, derived for special cases of Gaussian and MIMO channels, appear to be an instance of the general framework considered in the present work. This can be seen by assuming a *non-ergodic* channel model $\{W_\theta(y|x)\}$ (there are no time-varying states $s$), controlled by an unknown state $\theta \in \Theta$, where an estimate $\hat{\theta}$ of $\theta$ and its accuracy statistic $\mu_{\boldsymbol{\theta}|\hat{\boldsymbol{\theta}}}$ are available at the transmitter and at the receiver. We can consider the reliability function defined as the average of the transmission error probability over all channel estimation errors (this will be discussed later in Section II-C). It can be shown [19] that this notion of reliable communication leads to the capacity of the *composite channel*:

$$\mathbb{W}(y|x, \hat{\theta}) \doteq \int_\Theta W_\theta(y|x) d\mu(\theta|\hat{\theta}), \tag{2}$$

which results from the average of the unknown channel over the accuracy statistic, i.e., over all possible states given the estimate $\hat{\theta}$. The maximal achievable rate (the capacity) with reliable function defined by the average error probability over all channel estimation errors is

$$C(\hat{\theta}) = \sup_{P_{\hat{\theta}} \in \mathscr{P}_\Gamma(\mathscr{X})} I(X_{\hat{\theta}}; Y_{\hat{\theta}} | \hat{\boldsymbol{\theta}} = \hat{\theta}), \tag{3}$$

where $I(X_{\hat{\theta}}; Y_{\hat{\theta}} | \hat{\boldsymbol{\theta}} = \hat{\theta})$ is the mutual information evaluated for the composite channel (2) with input distribution $P_{\hat{\theta}}$ in the set of admissible input distributions $\mathscr{P}_\Gamma(\mathscr{X})$. Expression (3) represents the capacity for general DMCs with arbitrary estimation functions. For instance, the bounds for Gaussian inputs and MMSE channel estimation found in [15] and [18] can be derived as well from (3). However, note that Gaussian inputs are not optimal for maximizing this capacity and that only lower and upper bounds are known. The proof of (3) directly follows from Shannon's coding theorem [20], since the resulting error probability function can be defined in terms of the composite channel [21]. Moreover, it was shown in [19] that the capacity in (3) can be achieved by a ML decoder matched to the composite channel.

Consider now the case of *ergodic* models $\{W(y|x,s)\}$ controlled by an unknown sequence of states $\{s_i\}_{i=1}^{\infty}$ (there is no fixed and unknown state $\theta$), where the transmitter and the receiver are provided with the sequences $\{u_i\}_{i=1}^{\infty}$ and $\{v_i\}_{i=1}^{\infty}$, respectively, and these sequences are drawn i.i.d. from the joint PD $\mu_{\boldsymbol{SUV}}$. The results by Salehi [22], based again on the average error probability, extend Shannon's result [23] by showing that the capacity in this case is

$$C = \sup_{P_T \in \mathscr{P}_\Gamma} I(T;Y|V), \qquad (4)$$

where $T \in \mathscr{X}^{\|\mathscr{U}\|}$ is a random vector of length $\|\mathscr{U}\|$ with elements in $\mathscr{X}$ and $P_T$ is its PD. It is appropriate to mention here, that from the results of [9] the problem of imperfect CSI for *ergodic* channels with time-varying states is overcome by coding over expanded alphabets, where the estimates known at the receiver are considered as an additional output $(y,v)$ and those known at the transmitter as an additional input $(x,u)$. This simple argument stated in [9] shows that (4) follows again from Shannon's result [23], so that no proof is actually needed.

The capacity notions as defined above, based on averaging the reliability function over all channel estimation errors, cannot guarantee reliable communication in non-*ergodic* scenarios, specifically when significant differences arise between the true state $\theta$ and its estimate. In other words, the above notions consider a transmission successful if the "average" (over the ensemble of states $\theta$ given the estimate) of the error probability is small. This is therefore not really compatible with the constraints that are usually employed in *non-ergodic* scenarios, in which one wants to characterize the capacity attained by all users who have access to the service. In contrast, the notion of EIO capacity that we shall propose is closely related to that of outage capacity, originally introduced in [24]. It is also connected to the notion of $\epsilon$-capacity, first proved for a class of discrete stationary channels called regular decomposable (cf. [14], [25], [26]). In the standard scenario of slowly fading AWGN channels, where the receiver is provided with perfect CSI and there is not CSIT, this notion relies on the fact that there may be a non-negligible probability that the value of the actual transmitted rate exceeds the instantaneous mutual information and thus an outage event occurs. Hence the error probability does not decay when the block-length increases. The capacity versus outage is then defined as the maximum rate that can be supported with probability $1 - \gamma$, where $\gamma$ is a prescribed outage probability used to exclude the outage events. Indeed, it has been shown that the outage probability matches well the error probability of practical codes (cf. [27], [28]). The general results by Verdú and Han [29] provide a coding theorem for arbitrary channels in this setting. The $\epsilon$-capacity $C_\epsilon$ ($0 < \epsilon = \gamma < 1$) is given by [29]

$$C_\epsilon = \sup_{P_X^n \in \mathscr{P}_\Gamma(\mathscr{X}^n)} \sup \left\{ R \geq 0 : F_X(R, P_X^n) \leq \epsilon \right\}, \tag{5}$$

with the limit of *cumulative distribution functions* defined as follows

$$F_X(R, P_X^n) \doteq \limsup_{n \to \infty} \Pr \left( \frac{1}{n} \log \frac{W_\theta^n(Y_{\theta,1}, \ldots, Y_{\theta,n}, \boldsymbol{\theta} = \theta | X_1, \ldots, X_n)}{W_\theta^n P_X^n(Y_{\theta,1}, \ldots, Y_{\theta,n}, \boldsymbol{\theta} = \theta)} \leq R \right), \tag{6}$$

where $W_\theta^n$ and $P_X^n$ are the $n$-extensions of the channel and its input process. Note that the state $\theta$ in (6) is considered as an additional channel output [30]. Moreover, the general expression (5) holds also with imperfect CSIR by evaluating (6) with the channel (2), averaged over all state estimation errors, and letting the state estimate $\hat{\theta}$ instead of $\theta$ be an additional channel output.

In *non-ergodic* scenarios, a transceiver using $(\hat{\theta}, \{u_i\}_{i=1}^\infty, \{v_i\}_{i=1}^\infty)$ instead of $(\theta, \{s_i\}_{i=1}^\infty)$ obviously might not support a desired information rate, even arbitrarily small rates might not be supported if $\hat{\theta}$ and $\theta$ happen to be strongly different. As a consequence of this observation,

"outages" induced by channel estimation errors will occur with a certain probability. In this paper, we introduce the notion of *estimation-induced outage (EIO) capacity* that is a function of the outage probability $\gamma$ (a QoS parameter), the specific channel measurement $\hat{\theta}$ and the joint accuracy statistic $\mu_{\boldsymbol{\theta}SUV|\hat{\boldsymbol{\theta}}}$. A single-letter characterization, evaluating the optimal trade-off between the maximal EIO rate versus the outage probability, will be derived.

## C. Outline

The remainder of this paper is organized as follows. In Section II, the notion of EIO capacity is first formalized for general DMCs and then a coding theorem is stated. Section III presents the main steps of the proof of the coding theorem and its converse. An application example of a *non-ergodic* fading Ricean channel with ML channel estimation is considered in Section IV. The mean EIO capacity is compared to the achievable EIO rates of a system using the mismatched ML decoder based on the state estimate. The effects of quantized feedback and power allocation strategies are also considered. Section V provides numerical results to illustrate mean EIO rates.

## II. ESTIMATION-INDUCED OUTAGE CAPACITY AND CODING THEOREM

In this section, we first develop a proper formalization of the notion of *EIO capacity* and state a coding theorem.

## A. Notation

Throughout the next sections we use the following notation: $\mathscr{P}(\mathscr{X})$ denotes the set of all atomic (or discrete) PDs on $\mathscr{X}$ with finite number of atoms. Then the $n$-th Cartesian power is defined as the sample space of $\mathbf{X} = (X_1, \ldots, X_n)$, with $P_{\mathbf{X}}^n$-probability mass determined in terms of the $n$-th Cartesian power of $P_X$. The joint PD corresponding to the input $P_X \in \mathscr{P}(\mathscr{X})$ and the transition PD $W_\theta \in \mathscr{P}(\mathscr{Y})$, is denoted as $W_\theta \circ P_X \in \mathscr{P}(\mathscr{X} \times \mathscr{Y})$ and its marginal on $\mathscr{Y}$ is denoted as $W_\theta P_X \in \mathscr{P}(\mathscr{Y})$. The cardinality of the alphabets is denoted by $\|\cdot\|$, and the complement of any set $\mathscr{A}$ is denoted by $\mathscr{A}^c$, while $\mathbb{1}_{\{\cdot\}}$ denotes the indicator function. The functionals $\mathcal{D}(\cdot\|\cdot)$ and $H(\cdot)$ denote the *Kullback-Leibler divergence* and the entropy, respectively.

## B. Problem Definition

A message $m$ from the set $\mathcal{M}_n = \{1, \ldots, \lfloor\exp(nR_{\hat{\theta}})\rfloor\}$ is transmitted using a length-$n$ block code defined by a sequence of encoding functions $\varphi_{\hat{\theta}}^n \doteq \{\varphi_{\hat{\theta},i} : \mathcal{M}_n \times \Theta \times \mathscr{U}^i \mapsto \mathscr{X}\}_{i=1}^n$ provided with states $(\hat{\theta}, u_1, \ldots, u_i) \in \Theta \times \mathscr{U}^i$; the receiver uses a sequence of decoding functions $\phi_{\hat{\theta}}^n \doteq \{\phi_{\hat{\theta},i} : \mathscr{Y}^n \times \Theta \times \mathscr{V}^i \mapsto \mathcal{M}_n \cup \{0\}\}_{i=1}^n$ provided with states $(\hat{\theta}, v_1, \ldots, v_i) \in \Theta \times \mathscr{V}^i$. The maximum (over all messages) of the average (with respect to $\mathbf{s}$, $\mathbf{u}$, $\mathbf{v}$, $\mathbf{y}$) error probability, which depends on the unknown state $\theta$, is defined as

$$\bar{e}_{\max}^{(n)}(\varphi_{\hat{\theta}}^n, \phi_{\hat{\theta}}^n | \theta, \hat{\theta}) \doteq \max_{m \in \mathcal{M}_n} \sum_{\mathbf{s} \in \mathscr{S}^n} \sum_{\mathbf{u} \in \mathscr{U}^n} \sum_{\mathbf{v} \in \mathscr{V}^n} \sum_{\mathbf{y} \in \mathscr{Y}^n} \mathbb{1}_{\{\phi_{\hat{\theta}}^n(\mathbf{y}, \mathbf{v}) \neq m\}} W_\theta^n(\mathbf{y} | \varphi_{\hat{\theta}}^n(m, \mathbf{u}), \mathbf{s}) \mu^n(\mathbf{s}, \mathbf{u}, \mathbf{v} | \theta, \hat{\theta}). \tag{7}$$

Each transmitted codeword must satisfy a transmission cost constraint (generalized power constraint) $\mathbb{E}_{\mathbf{XU}}\{\Phi_{\hat{\theta}}(\mathbf{x}, \mathbf{u})\} \leq n\Gamma$ where $\Phi_{\hat{\theta}}(\mathbf{x}, \mathbf{u}) = \sum_{i=1}^n \Phi_{\hat{\theta}}(x_i, u_i)$ ($\Gamma \in \mathbb{R}_+$) for some cost function $\Phi_{\hat{\theta}} : \mathscr{X} \times \Theta \times \mathscr{U} \longmapsto \mathbb{R}_+$. In the absence of a transmission cost constraint, we set $\Gamma = \infty$.

*Definition 2.1:* For a given estimate $\hat{\theta}$ and $0 \leq \epsilon, \gamma < 1$, an EIO rate $R_{\hat{\theta}} \geq 0$ is said $(\epsilon, \gamma)$-achievable on a DMC $\{W_\theta(y|x,s)\}$, if for every $\delta > 0$ and sufficiently large $n$, there exists for the unknown state $\theta$ a block code of length-$n$ and size $M_{\theta\hat{\theta}}$ that supports an error probability (7) smaller than $\epsilon$ with prescribed outage probability:

$$\Pr\left(\{\theta \in \Lambda_\epsilon^{(n)} : n^{-1} \log M_{\theta\hat{\theta}} \geq R_{\hat{\theta}} - \delta\} \big| \hat{\boldsymbol{\theta}} = \hat{\theta}\right) \geq 1 - \gamma, \tag{8}$$

where $\Lambda_\epsilon^{(n)} = \{\theta \in \Theta \colon \bar{e}_{\max}^{(n)}(\varphi_{\hat{\theta}}^n, \phi_{\hat{\theta}}^n | \theta, \hat{\theta}) \leq \epsilon\}$ is the set of all channel states allowing for reliable decoding. In other words, the encoder uses a smaller codebook that guarantees maximum error probabilities less than $\epsilon$ with probability at least $1 - \gamma$.

A rate $R_{\hat{\theta}} \geq 0$ is $\gamma$-achievable if it is $(\epsilon, \gamma)$-achievable for every $0 < \epsilon < 1$. Let $C_{EIO}^{(\epsilon)}$ be the largest $(\epsilon, \gamma)$-achievable rate for an outage probability $\gamma$ and an estimate $\hat{\theta}$. The EIO capacity is then defined as the largest $\gamma$-achievable rate with $\epsilon \to 0$,

$$C_{EIO}(\gamma, \hat{\theta}) \doteq \lim_{\epsilon \downarrow 0} C_{EIO}^{(\epsilon)}(\gamma, \hat{\theta}).$$

We next compare the notion of reliable communication underlying the EIO capacity to the different reliability notions discussed in the introduction section.

(i) The practical advantage of Definition 2.1 is that, for each transmission with an unknown but fixed draw of $\theta$, the transmitter and the receiver are designed for guaranteeing maximum transmission rate for most of states, but the worst ones are considered as "outages" with probability $\gamma$. This provides more precise control over the reliability function (7) at the expense of decreasing the information rate. Notice that the conventional capacity [15] discussed in Section I, which only requires small "averaged" error probabilities

$$\bar{e}_{\max}^{(n)}(\varphi_{\hat{\theta}}^n, \phi_{\hat{\theta}}^n | \hat{\theta}) = \mathbb{E}_{\boldsymbol{\theta}|\hat{\boldsymbol{\theta}}}\{\bar{e}_{\max}^{(n)}(\varphi_{\hat{\theta}}^n, \phi_{\hat{\theta}}^n | \boldsymbol{\theta}, \hat{\theta}) | \hat{\boldsymbol{\theta}} = \hat{\theta}\} \leq \epsilon, \qquad (9)$$

does not guarantee small error probabilities for each transmission over a channel with an unknown state $\theta$ (or *non-ergodic* component). In contrast, with EIO capacity the encoder and decoder determine the most likely set of states $\theta$ using the available CSI $(\mathcal{W}_\Theta, \mu_{\boldsymbol{\theta SUV}|\hat{\boldsymbol{\theta}}}, \hat{\theta})$ and construct codes that perform well simultaneously for all states (channels) in that set. Hence, similarly to the notion of $\epsilon$-capacity (5), this approach requires to eliminate the worst (unlikely but possible) states since these would yield zero capacity values. In contrast, similarly to the notion of ergodic capacity [9], one can average the reliability function (7) over all channel estimation errors corresponding to the time-varying states $\{s_i\}_{i=1}^\infty$ (or *ergodic* components). Related problems regarding the capacity of compound and average quantum channels were reported in [31].

(ii) In the conventional definition of capacity versus outage when the coding rate chosen is greater than the instantaneous mutual information an outage event occurs. For instance, the mutual information specifies the maximum rate with error-free communication[2]. This definition implicitly assumes that the state $\theta$ is available at the receiver as an additional output and therefore cannot be directly extended in presence of $\hat{\theta}$. Indeed, notice that error-free communications cannot be guaranteed with this setting even for the best realizations of $\theta$. Imperfect CSIR with no CSIT can be directly considered via the $\epsilon$-capacity. To this end, one can average the channel over all state estimation errors, which would yield the channel model (2) with additional channel output $\hat{\theta}$, and then evaluate the general expression (5). In contrast, the EIO capacity allows for imperfect CSIT and roughly speaking, it is the maximal coding rate guaranteeing error-free communications for $(1-\gamma)$ percent of the states $\theta$ given an estimate, according to the statistics of estimation errors.

(iii) Assume that $\theta$ is available at the receiver (an additional output) and for simplicity suppose that there are no time-varying states ($U = V = S = \emptyset$). When $\theta$ is independent of $\hat{\theta}$, we observe that the functional (8) reduces to the conventional $\epsilon$-capacity (5) (except for a set of states $\theta$ with zero measure [29]). This can be easily seen by noting that in this case, $\Lambda_\epsilon^{(n)} = \Theta$ and the rate of the code coincides with the (instantaneous) mutual information. Hence, expression (8) becomes $\Pr\left(\{\theta \in \Theta : I(X; Y_\theta, \boldsymbol{\theta} = \theta) \geq R - \delta\}\right) \geq 1 - \gamma$ that equals (6) for memoryless sequences $\{X_n\}_{n=1}^\infty$. The $\gamma$-capacity follows by taking the supremum over all rates $R \geq 0$.

---

[2]Here, error-free communications is understood in the sense of arbitrarily small error probabilities in the limit.

## C. Coding Theorem

The next theorem quantifies the EIO capacity and provides an explicit way to evaluate the maximal EIO rate versus the outage probability $\gamma$ for an arbitrary DMC controlled by a random state sequence $(\theta, \{s_i\}_{i=1}^{\infty})$. The transmitter and the receiver are provided with the sequence of state estimates $(\hat{\theta}, \{u_i\}_{i=1}^{\infty})$ and $(\hat{\theta}, \{v_i\}_{i=1}^{\infty})$ and the joint statistic $\mu_{\theta SUV|\hat{\theta}}$, respectively.

*Theorem 2.2 (EIO capacity):* Given an outage probability $0 \leq \gamma < 1$ and sequences of state estimates, the EIO capacity of an arbitrary DMC $\{W_\theta(y|x,s)\}$ is given by

$$C_{EIO}(\gamma, \hat{\theta}) = \sup_{q_{TX|U\hat{\theta}} \in \mathscr{P}_\Gamma} \sup_{\Lambda \subset \Theta_\gamma} \inf_{\theta \in \Lambda} I(T; Y_\theta | V_\theta, \hat{\boldsymbol{\theta}} = \hat{\theta}), \tag{10}$$

where the set of admissible input PDs is defined as

$$\mathscr{P}_\Gamma \doteq \Big\{ q_{TX|U\hat{\theta}} \in \mathscr{P}(\mathscr{T} \times \mathscr{X}) : q_{TX|U\hat{\theta}} = \mathbb{1}_{\{X = f_T(\hat{\theta}, U)\}} P_{T|\hat{\theta}},$$

$$T \multimap (X, S) \multimap Y_\theta \ \forall\, \theta \in \Theta,\ \mathscr{T} = \mathscr{X}^{\|\mathscr{U}\|},\ \mathbb{E}_{XU}\{\Phi_{\hat{\theta}}(X, U)\} \leq \Gamma \Big\}$$

with mappings $\{f_t : \Theta \times \mathscr{U} \longmapsto \mathscr{X}\}_{t \in \mathscr{T}}$ and $\Theta_\gamma \doteq \{\Lambda \subseteq \Theta : \Pr(\Lambda | \hat{\boldsymbol{\theta}} = \hat{\theta}) \geq 1 - \gamma\}$. The supremum in (10) is taken over all subsets $\Lambda \subseteq \Theta$ that have (conditional) probability at least $1 - \gamma$ and the mutual information is given by

$$I(T; Y_\theta | V_\theta, \hat{\boldsymbol{\theta}} = \hat{\theta}) = \sum_{t \in \mathscr{T}} \sum_{y \in \mathscr{Y}} \sum_{v \in \mathscr{V}} P_{T|\hat{\theta}}(t|\hat{\theta}) \mathbb{W}_\theta(y, v|t, \hat{\theta}) \log \frac{\mathbb{W}_\theta(y, v|t, \hat{\theta})}{\mathbb{W}_\theta(y, v|\hat{\theta})}, \tag{11}$$

where

$$\mathbb{W}_\theta(y, v|t, \hat{\theta}) = \sum_{x \in \mathscr{X}} \sum_{u \in \mathscr{U}} \mathbb{W}_\theta(y, v|x, u) \mathbb{1}_{\{x = f(t, \hat{\theta}, u)\}} \mu(u|\theta, \hat{\theta}) \tag{12}$$

and the equivalent channel with inputs $(x, u, \hat{\theta})$ and outputs $(y, v, \hat{\theta})$ is given by

$$\mathbb{W}_\theta(y, v|x, u, \hat{\theta}) = \sum_{s \in \mathscr{S}} W_\theta(y|x, s) \mu(s, v|u, \theta, \hat{\theta}). \tag{13}$$

*Comments:* (i) The expression of the EIO capacity in Theorem 2.2 provides a general formula for arbitrary state-dependent DMCs with imperfect CSI.

Using basic information-theoretic considerations, it can be seen that the capacities and reliability measures discussed in the introduction are special cases of the EIO capacity. Therefore, the EIO capacity can be viewed as a unification of the results in [9], developed for *ergodic* channels with imperfect CSIT and CSIR, and the natural extension of the results in [29], originally derived for general channels with no CSIT. We mention at this point that (10) can be reached from the $\epsilon$-capacity (5) by letting the receiver known $(\theta, \hat{\theta}, v)$ while the transmitter observes $(\hat{\theta}, u)$. Furthermore, if the transmitter has a noisy version $\tilde{\theta}$ of $\hat{\theta}$ (e.g. due to quantization and/or feedback errors) while the receiver is aware of $(\tilde{\theta}, \hat{\theta})$, Theorem 2.2 still holds with $\hat{\theta}$ in (10) replaced by $\tilde{\theta}$.

(ii) The goal of the encoder and the decoder in the EIO capacity is to determine the set of states $\Lambda^\star = \{\theta \in \Theta\}$ that maximize the information rate over the averaged channel (12) and simultaneously have sufficiently high probability given the state estimate $\hat{\theta}$. Then the encoder constructs codes that perform well simultaneously over all channel states $\theta \in \Lambda^\star$. Hence, it should be noted that this approach yields a compound setting [14], [32] of the averaged channel investigated in [22]. This point of view can be complemented by the observation that compound channels play the role of the simplest models for situations where channel uncertainty arises in the *non-ergodic* components (the state $\theta$) of the channel statistic controlling the communication. While averaged channels model scenarios in which the uncertainty is present in the *ergodic* components (the time-varying states $s$) of the channel. Furthermore, from (10) we can observe that the EIO capacity is not increased if the receiver is informed with the true channel state $\theta$, but not the encoder. This observation coincides as well with the capacity results for conventional compound channels [1], [21] and quantum compound channels [33], [34].

(iii) The proof of Theorem 2.2 is based on bounding the minimum size of the image of a code through a channel. The details of the proof and some associated technical aspects are relegated to Section III and Appendix II, respectively. Although there exist alternative ways of proving this theorem, e.g. by using *universal decoders* (cf. [6], [7]), the present proof illustrates the connections between EIO capacity and the capacity of conventional compound channels and the manner in which the available CSI is exploited. Furthermore, it may perhaps provide useful insights regarding practical code design. The generalized *Maximal Code Lemma* used in our proof can be extended to more general models, as for example *mismatched decoders*.

The generalization of the theorem and its proof to continuous alphabets is complicated by the fact that continuous-alphabet extensions of the concept of *types* (which is used in our proof) are not known [35]. Yet, there are several continuous-alphabet problems whose simplest (or only) solution relies upon the method of types via discrete approximations. The proof of Sanov's theorem in [36] and the capacity of Arbitrarily Varying Channels (AVC) with general alphabets and states have been determined in this way (cf. [37]). A possible route for a generalization of Theorem 2.2 to continuous alphabets is the use of the weak topology, requiring different tools from measure theory and consideration of locally compact Hausdorff spaces, e.g. alphabets like $\mathbb{R}^k$ (or $\mathbb{C}^k$) which are separable spaces. However, this extension is not considered in this paper.

### D. Impact of Channel Estimation Errors on the EIO Capacity

We now present a general upper bound on the rate loss with respect to the perfect CSI scenario. To this end, we upper bound the rate difference between the EIO capacity (10) and the *ergodic* capacity with perfect CSI. Notice that we compare to the *ergodic* capacity because with high-accuracy estimations $\mu_{\boldsymbol{\theta}|\hat{\boldsymbol{\theta}}}$ becomes close to a Dirac distribution and thus the EIO capacity approaches (with probability $\gamma$ close to one) the ergodic capacity. The following Lemma easily follows as a consequence of Theorem 3.1, stated and proved in Appendix III.

*Lemma 2.3:* Assume that the optimal set of channel states $\Lambda^\star \subset \Theta$ obtained by maximizing the EIO capacity (10) (over all sets of states $\Lambda \subseteq \Theta$ having probability at least $1 - \gamma$) defines a convex set of conditional PDs $\mathcal{W}_{\Lambda^\star} \doteq \{ \mathbb{W}_\theta : \mathcal{T} \times \Theta \longmapsto \mathcal{Y} \times \mathcal{V} \}_{\theta \in \Lambda^\star}$. Let $\theta^\star \in \Lambda^\star$ be the channel state that provides the infimum in expression (10). The following inequality holds

$$C_{EIO}(\gamma, \hat{\theta}) \leq C_{\mathrm{E}}(\theta) - \left[ \mathcal{D}\big((Y_\theta, V_\theta) \| (Y_{\theta^\star}, V_{\theta^\star}) | T, \hat{\theta}\big) - \mathcal{D}\big((Y_\theta, V_\theta) \| (Y_{\theta^\star}, V_{\theta^\star}) | \hat{\theta}\big) \right], \qquad (14)$$

for any arbitrary state $\theta \in \Lambda^\star$ and the corresponding input $T$ with PD $q_{TX|U\hat{\theta}} \in \mathscr{P}_\Gamma$ that maximizes the expression of the EIO capacity in (10), where $C_{\mathrm{E}}(\theta)$ denotes the ergodic capacity

$$C_{\mathrm{E}}(\theta) = \sup_{q_{X|\theta} \in \mathscr{P}_\Gamma} I(X; Y_\theta | S_\theta).$$

Notice that the term in brackets on the right-hand side of the inequality (14) is positive. Moreover, equality in (14) holds for all *linear families* of conditional PDs (or channels) $\mathcal{W}_{\Lambda^\star}$.

### III. Proof of the Coding Theorem and Its Converse

In this section we approach the problem of determining optimal codes for achieving the EIO capacity, according to its definition in Section II-B. The proof of Theorem 2.2 is based on a generalization of the *Maximal Code Lemma* [38] to bound the minimum size of the image of a code through the considered class of DMCs. Roughly speaking, the encoder by using the available information determines the most likely set of channel states and constructs codes that perform well for all states in this set. Decoding is based on the union of I-typical sets, which are called *robust I-typical sets* (see Appendix II).

## A. Generalized Maximal Code Lemma

This section uses the notion of information-typical (I-typical) and conditionally I-typical sets defined in terms of the Kullback-Leibler divergence: $\mathcal{T}^n_{[T]_\delta} = \{\mathbf{t} \in \mathcal{T}^n : \mathcal{D}(\hat{P}_n\|P_T) \leq \delta\}$ and $\mathcal{T}^n_{[Y_\theta V_\theta|T]_\delta}(\mathbf{t}) = \{(\mathbf{y}, \mathbf{v}) \in \mathcal{Y}^n \times \mathcal{V}^n : \mathcal{D}(\widehat{W}_n\|\mathbb{W}_\theta|\hat{P}_n) \leq \delta\}$ (for further details see Appendix I). Furthermore, we define the set of states

$$\Lambda_\epsilon \doteq \Big\{\theta \in \Theta : \min_{(\mathbf{x},\mathbf{u}) \in \mathcal{T}^n_{[XU|T\hat{\theta}]_\delta}(\mathbf{t}_m)} \mathbb{W}^n_\theta\big(\mathscr{D}^n_m|\mathbf{x}, \mathbf{u}, \hat{\theta}\big) > 1 - \epsilon, \text{ for all } m \in \mathcal{M}_n\Big\}, \qquad (15)$$

such that $\Pr(\Lambda_\epsilon|\hat{\boldsymbol{\theta}} = \hat{\theta}) \geq 1 - \gamma$.

*Definition 3.1 (Admissible code):* Given a state sequence $(\theta, \{s_i\}_{i=1}^\infty)$, the encoder and the decoder are provided with the sequence of state estimates $(\hat{\theta}, \{u_i\}_{i=1}^\infty)$ and $(\hat{\theta}, \{v_i\}_{i=1}^\infty)$, respectively. The decoder reads $\phi_{\hat{\theta}}(\mathbf{y}, \mathbf{v}) = m$ iff $m$ is the only message such that $(\mathbf{y}, \mathbf{v}) \in \mathscr{D}^n_m$ ($\mathscr{D}^n_m$ denotes the decoding set associated to message $m$ [38]), while the encoder sends $\mathbf{x} = \varphi_{\hat{\theta}}(m, \mathbf{u}) = f^n_{\mathbf{t}_m}(\hat{\theta}, \mathbf{u})$. For an arbitrary set $\mathscr{T}_0 \subset \mathcal{T}^n \cap \mathcal{T}^n_{[T|\hat{\theta}]_\delta}$ with $P^n_{T|\hat{\theta}}(\mathscr{T}_0) \geq \eta$, $0 < \delta, \epsilon, \eta, \gamma < 1$ and an input PD $\{q_{TX|U\hat{\theta}} \in \mathscr{P}_\Gamma\}$ with mappings $\{f_t : \Theta \times \mathcal{U} \longmapsto \mathcal{X}\}_{t \in \mathcal{T}}$, an admissible $(n, \epsilon)$-code has to satisfy the following requirements:

(i) all codewords, depending only on the estimate $\hat{\theta}$, satisfy $\{\mathbf{t}_m\}_{m \in \mathcal{M}_n} \subset \mathscr{T}_0$;

(ii) all decoding sets $\{\mathscr{D}^n_m\}_{m \in \mathcal{M}_n} \subset \mathcal{Y}^n \times \mathcal{V}^n$ (depending on $\hat{\theta}$) are mutually disjoint;

(iii) for every $m \in \mathcal{M}_n$, the decoding sets satisfy

$$\mathscr{D}^n_m \subset \mathcal{T}^n_{[Y_\Lambda V_\Lambda|T\hat{\theta}]_\delta}(\mathbf{t}_m, \hat{\theta}) \doteq \bigcup_{\theta \in \Lambda_\epsilon} \mathcal{T}^n_{[Y_\theta V_\theta|T\hat{\theta}]_\delta}(\mathbf{t}_m, \hat{\theta}).$$

We now state a *Fundamental Lemma* analogue to Feinstein's Lemma [39] and its converse that bound the size of any code through the considered time-varying DMCs with imperfect CSI. The proof of Theorem 2.2 is immediate from this Lemma.

19*Lemma 3.2 (Fundamental Lemma):* For every $0 < \epsilon, \delta, \eta, \tau, \gamma < 1$ and channel estimate $\hat{\theta}$, every DMC $\{\mathbb{W}_\theta : \mathscr{X} \times \mathscr{U} \times \Theta \longmapsto \mathscr{Y} \times \mathscr{V}\}$ and admissible input PD $\{q_{TX|U\hat{\theta}} \in \mathscr{P}_\Gamma\}$, and every set $\mathscr{T}_0 \subset \mathscr{T}^n$ satisfying $P^n_{T|\hat{\theta}}(\mathscr{T}_0) \geq \eta$, there exists a positive integer $n_0(\|\mathscr{S}\|, \|\mathscr{U}\|, \|\mathscr{T}\|, \|\mathscr{X}\|, \|\mathscr{Y}\|, \epsilon, \delta, \eta)$ such that for all $n \geq n_0$ the following statements hold.

1) *Direct part:* There exists a set of states $\Lambda^\star \subset \Theta$ and admissible $(n, \epsilon)$-codes with codeword set $\{\mathbf{t}_m\}_{m \in \mathcal{M}_n} \subset \mathscr{T}_0 \cap \mathcal{T}^n_{[T|\hat{\theta}]_\delta}$ and rate $n^{-1} \log M_{\theta\hat{\theta}}$, whose maximum error probability (7) is smaller than $\epsilon$ for all $\theta \in \Lambda^\star$ and such that

$$\Pr\left(\{\theta \in \Lambda^\star : n^{-1} \log M_{\theta\hat{\theta}} \geq R_{\hat{\theta}} - 2\delta\} \big| \hat{\boldsymbol{\theta}} = \hat{\theta}\right) \geq 1 - \gamma, \tag{16}$$

for all rates $R_{\hat{\theta}} \leq C_{EIO}(\gamma, \hat{\theta}, q_{TX|U\hat{\theta}})$.

2) *Converse part:* For any admissible $(n, \epsilon)$-code of rate $n^{-1} \log M_{\theta\hat{\theta}}$, whose maximum error probability is smaller than $\epsilon$ for every $\theta \in \Lambda_\epsilon$, the largest code size satisfies

$$\Pr\left(\{\theta \in \Lambda_\epsilon : n^{-1} \log M_{\theta\hat{\theta}} > R_{\hat{\theta}} + 2\delta\} \big| \hat{\boldsymbol{\theta}} = \hat{\theta}\right) < \gamma, \tag{17}$$

for all rates $R_{\hat{\theta}} \geq C_{EIO}(\gamma, \hat{\theta}, q_{TX|U\hat{\theta}})$.

The proof of Lemma 3.2 is obtained from basic properties of common $\eta$-images and the concept of robust I-typical sets developed in Appendix II.

*Definition 3.3 (Common images of sets via channels):* A set $\mathscr{B} \subset \mathscr{Y}^n \times \mathscr{V}^n$ is a common $\eta$-image ($0 < \eta \leq 1$) of a set $\mathscr{T}_0 \subset \mathscr{T}^n$ via the collection of simultaneous DMCs $\{\mathbb{W}_\theta : \mathscr{T} \times \Theta \longmapsto \mathscr{Y} \times \mathscr{V}\}_{\theta \in \Lambda}$ if $\mathbb{W}^n_\theta(\mathscr{B}|\mathbf{t}, \hat{\theta}) \geq \eta$ for all $\theta \in \Lambda$ and every $\mathbf{t} \in \mathscr{T}_0$. Thus, the set of all $\eta$-images is defined by

$$\mathscr{G}_{\mathbb{W}_\Lambda}(\mathscr{T}_0, \hat{\theta}, \eta) \doteq \left\{\mathscr{B} \subset \mathscr{Y}^n \times \mathscr{V}^n : \inf_{\theta \in \Lambda} \mathbb{W}^n_\theta(\mathscr{B}|\mathbf{t}, \hat{\theta}) \geq \eta, \text{ for all } \mathbf{t} \in \mathscr{T}_0\right\}.$$

The minimum of the cardinalities of all common $\eta$-images $\mathscr{B}$ is denoted as

$$\mathrm{g}_{\mathbb{W}_\Lambda}(\mathscr{T}_0, \hat{\theta}, \eta) \doteq \min\left\{\|\mathscr{B}\| : \mathscr{B} \subseteq \mathscr{G}_{\mathbb{W}_\Lambda}(\mathscr{T}_0, \hat{\theta}, \eta)\right\}. \tag{18}$$

*Proof:* It is not difficult to show the existence of a subset $\mathcal{T}_0' \subset \mathcal{T}_0 \cap \mathcal{T}^n_{[T|\hat\theta]_\delta}$ such that $P^n_{T|\hat\theta}(\mathcal{T}_0') \geq \eta/2$ for sufficiently large $n$. Thus we can search for a codeword set $\{\mathbf{t}_m\}_{m\in\mathcal{M}_n} \subset \mathcal{T}_0'$. From Lemma 2.5 (Appendix II) we know that by choosing a confidence set $\Lambda \subset \Theta$ with $\Pr(\Lambda|\hat{\boldsymbol{\theta}} = \hat\theta) \geq 1-\gamma$ every robust I-typical set $\mathcal{T}^n_{[Y_\Lambda V_\Lambda|T\hat\theta]_\delta}(\mathbf{t}, \hat\theta) \subset \mathcal{Y}^n \times \mathcal{V}^n$ forms a robust $\epsilon$-decoding set (Definition 2.2) for each codeword $\mathbf{t} \in \mathcal{T}_0'$. Consider an admissible $(n, \epsilon)$-code that is maximal, which means that it cannot be extended by arbitrary $(\mathbf{t}_{M_{\theta\hat\theta}+1}; \mathscr{D}^n_{M_{\theta\hat\theta}+1})$ such that the extended code remains admissible. Define the set $\mathscr{D}^n \doteq \bigcup_{m=1}^{M_{\theta\hat\theta}} \mathscr{D}^n_m$ with $\mathscr{D}^n_m \subseteq \mathcal{T}^n_{[Y_\Lambda V_\Lambda|T\hat\theta]_\delta}(\mathbf{t}_m, \hat\theta)$, and choose $\tau < \epsilon$ such that $1-\epsilon > \epsilon - \tau$. It follows that

$$\inf_{\theta\in\Lambda}\left\{\min_{(\mathbf{x},\mathbf{u})\in\mathcal{T}^n_{[XU|T\hat\theta]_\delta}(\mathbf{t}_m)} \mathbb{W}^n_\theta(\mathscr{D}^n|\mathbf{x},\mathbf{u},\hat\theta)\right\} > \epsilon - \tau, \quad \text{for all } m \in \mathcal{M}_n. \tag{19}$$

For any $\mathbf{t} \in \mathcal{T}_0' \setminus \{\mathbf{t}_1, \ldots, \mathbf{t}_{M_{\theta\hat\theta}}\}$ and for all $\theta \in \Lambda$, if

$$\min_{(\mathbf{x},\mathbf{u})\in\mathcal{T}^n_{[XU|T\hat\theta]_\delta}(\mathbf{t})} \mathbb{W}^n_\theta\big(\mathcal{T}^n_{[Y_\Lambda V_\Lambda|T\hat\theta]_\delta} \setminus \mathscr{D}^n|\mathbf{x},\mathbf{u},\hat\theta\big) > 1 - \epsilon,$$

then the code would have an admissible extension, contradicting our initial assumption. Hence, for all $\mathbf{t} \in \mathcal{T}_0' \setminus \{\mathbf{t}_1, \ldots, \mathbf{t}_{M_{\theta\hat\theta}}\}$, we have

$$\inf_{\theta\in\Lambda}\left\{\max_{(\mathbf{x},\mathbf{u})\in\mathcal{T}^n_{[XU|T\hat\theta]_\delta}(\mathbf{t})} \mathbb{W}^n_\theta\big(\mathcal{T}^n_{[Y_\Lambda V_\Lambda|T\hat\theta]_\delta} \setminus \mathscr{D}^n|\mathbf{x},\mathbf{u},\hat\theta\big)\right\} \leq 1 - \epsilon.$$

The above expression and (19) imply for large enough $n$ that

$$\inf_{\theta\in\Lambda} \mathbb{W}^n_\theta(\mathscr{D}^n|\mathbf{t},\hat\theta) \geq (\epsilon-\tau)^2, \quad \text{for all } \mathbf{t} \in \mathcal{T}_0' \setminus \{\mathbf{t}_1, \ldots, \mathbf{t}_{M_{\theta\hat\theta}}\}. \tag{20}$$

Inequalities (19) and (20) actually imply that $\mathscr{D}^n$ is a common $(\epsilon-\tau)^2$-image of the set $\mathcal{T}_0'$ via the collection of DMCs $\{\mathbb{W}_\theta : \mathcal{T} \times \Theta \longmapsto \mathcal{Y} \times \mathcal{V}\}_{\theta\in\Lambda}$. By definition of $g_{\mathbb{W}_\Lambda}(\mathcal{T}_0', \hat\theta, (\epsilon-\tau)^2)$ it follows that

$$\|\mathscr{D}^n\| \geq g_{\mathbb{W}_\Lambda}(\mathcal{T}_0', \hat\theta, (\epsilon-\tau)^2). \tag{21}$$

On the other hand, since $\mathscr{D}_m^n \subseteq \mathscr{T}_{[Y_\Lambda V_\Lambda | T\hat{\theta}]_\delta}^n(\mathbf{t}_m, \hat{\theta})$ we have that

$$\|\mathscr{D}^n\| \leq \sum_{m=1}^{M_{\theta\hat{\theta}}} \|\mathscr{D}_m^n\| \leq M_{\theta\hat{\theta}} \max_{m \in \mathcal{M}_n} \|\mathscr{T}_{[Y_\Lambda V_\Lambda | T\hat{\theta}]_\delta}^n(\mathbf{t}_m, \hat{\theta})\|$$

$$\leq M_{\theta\hat{\theta}} \exp\left[n\left(\sup_{\theta \in \Lambda} H(Y_\theta, V_\theta | T_{\hat{\theta}}, \hat{\boldsymbol{\theta}} = \hat{\theta}) + \tau\right)\right], \quad (22)$$

for sufficiently large $n$ and all $\theta \in \Lambda$, where the last inequality follows by applying the upper bound of Theorem 2.6. Up to now by combining expressions (21) and (22), we have shown that there exists admissible $(n, \epsilon)$-codes such that

$$n^{-1} \log M_{\theta\hat{\theta}} \geq n^{-1} \log \mathrm{g}_{\mathbb{W}_\Lambda}\left(\mathscr{T}_0', \hat{\theta}, (\epsilon - \tau)^2\right) - \sup_{\theta \in \Lambda} H(Y_\theta, V_\theta | T_{\hat{\theta}}, \hat{\boldsymbol{\theta}} = \hat{\theta}) - \tau, \quad (23)$$

for all $\theta \in \Lambda$ and arbitrary set $\Lambda \subset \Theta$ having probability at least $1 - \gamma$. Let $\hat{\mathscr{D}}^n$ be the common $(\epsilon - \tau)^2$-image of minimal size $\|\hat{\mathscr{D}}^n\| = \mathrm{g}_{\mathbb{W}_\Lambda}\left(\mathscr{T}_0', \hat{\theta}, (\epsilon - \tau)^2\right)$. Then it can be seen that $\inf_{\theta \in \Lambda} \mathbb{W}_\theta P_{T|\hat{\theta}}^n(\hat{\mathscr{D}}^n) \geq \eta/2(\epsilon - \tau)^2$. By applying Lemma 1.7 (Appendix I) to this relation and substituting it in (23), we obtain

$$n^{-1} \log M_{\theta\hat{\theta}} \geq \inf_{\theta \in \Lambda} I\left(T_{\hat{\theta}}; Y_\theta, V_\theta | \hat{\boldsymbol{\theta}} = \hat{\theta}\right) - 2\tau, \quad (24)$$

for all $\theta \in \Lambda$ and $n \geq n_0'$, which follows by using the inequality $n^{-1} \log \mathrm{g}_{\mathbb{W}_\Lambda}\left(\mathscr{T}_0', \hat{\theta}, (\epsilon - \tau)^2\right) \geq H(Y_\theta, V_\theta | \hat{\boldsymbol{\theta}} = \hat{\theta}) - \tau$ for all $\theta \in \Lambda$. Finally, taking the supremum in (24) with respect to all sets $\Lambda \subset \Theta$ having probability at least $1 - \gamma$ yields the lower bound (16)

$$\begin{aligned} n^{-1} \log M_{\theta\hat{\theta}} &\geq C_{EIO}(\gamma, \hat{\theta}, q_{TX|U\hat{\theta}}) - 2\tau \\ &\geq R_{\hat{\theta}} - 2\tau, \end{aligned} \quad (25)$$

for all $R_{\hat{\theta}} \leq C_{EIO}(\gamma, \hat{\theta}, q_{TX|U\hat{\theta}})$ and states $\theta \in \Lambda^\star$, which is attained by setting $\Lambda_\epsilon = \Lambda^\star$.

We now prove the second statement (converse part). For every $\theta \in \Lambda_\epsilon$ and set $\Lambda_\epsilon \subset \theta$, let $\hat{\mathscr{D}}_\theta^n \subset \mathscr{Y}^n \times \mathscr{V}^n$ be an arbitrary set such that

$$\min_{(\mathbf{x},\mathbf{u}) \in \mathscr{T}_{[XU|T\hat{\theta}]_\delta}^n(\mathbf{t}_m)} \mathbb{W}_\theta^n\left(\hat{\mathscr{D}}_\theta^n | \mathbf{x}, \mathbf{u}, \hat{\theta}\right) \geq \epsilon + \tau, \quad \text{for every } m \in \mathcal{M}_n. \quad (26)$$

Hence, it follows from Definition (15) and expression (26) that

$$\mathbb{W}_\theta^n(\mathscr{D}_m^n \cap \hat{\mathscr{D}}_\theta^n | \mathbf{t}_m, \hat{\theta}) \geq \tau^2, \quad \text{for } m \in \mathcal{M}_n, \tag{27}$$

provided by $P_{XU|T\hat{\theta}}^n(\mathcal{T}_{[XU|T\hat{\theta}]_\delta}^n | \mathbf{t}_m) \geq \tau$ for sufficiently large $n$. Using Corollary 1.2.14 in [40], we hence obtain

$$\min_{m \in \mathcal{M}_n} \|\mathscr{D}_m^n \cap \hat{\mathscr{D}}_\theta^n\| \geq \exp\left[n\left(H(Y_\theta, V_\theta | T_{\hat{\theta}}, \hat{\boldsymbol{\theta}} = \hat{\theta}) - \tau\right)\right], \tag{28}$$

for each $\theta \in \Lambda_\epsilon$, provided $n$ is sufficiently large. Observe that we can set $\hat{\mathscr{D}}_\theta^n = \mathcal{T}_{[Y_\theta V_\theta | T\hat{\theta}]_\delta}^n(\mathbf{t}_m, \hat{\theta})$, which satisfies (26) for sufficiently large $n$ and every $m \in \mathcal{M}_n$. From the disjointness of decoding sets $\{\mathscr{D}_m^n\}_{m \in \mathcal{M}_n}$ of every admissible code, we have that

$$\|\hat{\mathscr{D}}_\theta^n\| \geq \sum_{m=1}^{M_{\theta\hat{\theta}}} \|\mathscr{D}_m^n \cap \hat{\mathscr{D}}_\theta^n\|$$

$$\geq M_{\theta\hat{\theta}} \exp\left[n\left(H(Y_\theta, V_\theta | T_{\hat{\theta}}, \hat{\boldsymbol{\theta}} = \hat{\theta}) - \tau\right)\right], \tag{29}$$

for all $\theta \in \Lambda_\epsilon$, where the last inequality follows from (28). Thus, we have shown that

$$n^{-1} \log M_{\theta\hat{\theta}} \leq n^{-1} \log \|\hat{\mathscr{D}}_\theta^n\| - H(Y_\theta, V_\theta | T_{\hat{\theta}}, \hat{\boldsymbol{\theta}} = \hat{\theta}) + \tau, \tag{30}$$

for all $\theta \in \Lambda_\epsilon$. Notice that since by assumption $\mathcal{T}_0 \subset \mathcal{T}_{[T|\hat{\theta}]_\delta}^n$, it follows that $\hat{\mathscr{D}}_\theta^n \subset \mathcal{T}_{[Y_\theta V_\theta | \hat{\theta}]_\delta}^n$ and thus Proposition 1.4-(iv) (see Appendix I) shows that there exists $n \geq n_0''$ such that

$$n^{-1} \log \|\hat{\mathscr{D}}_\theta^n\| \leq H(Y_\theta, V_\theta | \hat{\boldsymbol{\theta}} = \hat{\theta}) + \tau. \tag{31}$$

Hence, by applying (31) to (30) and then taking its supremum with respect to all sets $\Lambda \subset \Theta$ having probability at least $1 - \gamma$, we obtain that for all $\theta \in \Lambda$

$$\begin{aligned} n^{-1} \log M_{\theta\hat{\theta}} &\leq C_{EIO}(\gamma, \hat{\theta}, q_{TX|U\hat{\theta}}) + 2\tau, \\ &\leq R_{\hat{\theta}} + 2\delta, \end{aligned} \tag{32}$$

with $R_{\hat{\theta}} \geq C_{EIO}(\gamma, \hat{\theta}, q_{TX|U\hat{\theta}})$ and $\Pr(\boldsymbol{\theta} \notin \Lambda | \hat{\theta}) < \gamma$, which concludes the proof. ∎

# IV. APPLICATION EXAMPLE: EIO CAPACITY OF THE NON-ERGODIC RICIAN FADING CHANNEL

In this section, we illustrate the results by evaluating the EIO capacity using Theorem 2.2. We consider a simple but rich enough framework that assumes communication over a single-antenna wireless channel, involving a Rician block flat fading model, where the channel state $\theta \in \Theta$ is described by a single coefficient $\boldsymbol{\theta} = H \in \mathbb{C}$ and it is assumed that there are no time-varying states ($U = V = S = \emptyset$). The single channel state $\theta$ is assumed to remain constant during the transmission of each codeword but it is unknown at the transmitter and the receiver. Each transmission is preceded by a short phase of channel training (which is small compared to the coherence time). This consists in sending a training sequence consisting of $N$ symbols, which are perfectly known at the receiver. Thus, the receiver is able to perform ML or MMSE estimation of $h$, yielding the noisy channel estimate $\hat{\theta} = \hat{h}$.

In many wireless systems, CSI at the transmitter is provided by the receiver via a feedback link, allowing the transmitter to use adaptive modulation and coding and to perform power control. We will consider the following three scenarios.

- No feedback channel is available (i.e. absence of CSIT).
- A perfect feedback link is available (i.e. the transmitter knows the actual estimate $\hat{\theta}$). For this case, we compare the EIO capacity with the EIO rates achievable with a receiver that performs mismatched ML decoding based on $\hat{\theta}$.
- A rate-limited feedback link is available, i.e., a quantized version $\tilde{\theta}$ of the estimate $\hat{\theta}$ is sent to the transmitter. The quantization codebook (designed using the well-known Lloyd-Max algorithm [41]) is known at both the transmitter and the receiver.

## A. Channel Model and Estimator Statistics

Consider a single-antenna block fading channel for wireless environments, given by

$$Y_i = HX_i + Z_i, \quad i = 1, \ldots, n, \tag{33}$$

where $Y_i \in \mathbb{C}$ is the received discrete-time signal, $X_i \in \mathbb{C}$ denotes the transmit signal, $H = h \in \Theta \doteq \mathbb{C}$ is the channel realization and $Z_i \in \mathbb{C}$ is the additive noise. Each transmitted codeword $\mathbf{x} = (x_1, \ldots, x_n)$ must satisfy the power constraint $\mathbb{E}_{\mathbf{X}}\{\|\mathbf{x}\|^2\} \leq n\Gamma_{\hat{\theta}}$ with power $\Gamma_{\hat{\theta}}$. The noise $Z_i$ is i.i.d. zero-mean circularly complex Gaussian (ZMCCG) with variance $\sigma_Z^2$. To model Rician fading, the channel state $\boldsymbol{\theta} = H$ is assumed to be circularly complex Gaussian with mean $\mu_H$ and variance $\sigma_H^2$, i.e., $H \sim C\mathcal{N}(\mu_H, \sigma_H^2)$. The Rice factor is defined as $K_H = \dfrac{|\mu_H|^2}{\sigma_H^2}$. The channel is a memoryless *non-ergodic* channel with conditional PD

$$W_H(y|x) = C\mathcal{N}(Hx, \sigma_Z^2) \quad \text{and} \quad \mathcal{W}_\Theta = \{W_{H=h}(y|x),\ h \in \Theta\}. \tag{34}$$

We are going to employ the EIO capacity expression provided by (10) with the appropriate transmission constraint, even though we provided a proof only for discrete input and output alphabets (see the comments at the end of Section II-C).

Since the channel coefficient $H = h$ is constant within a frame, channel estimation can be performed on the basis of known training (pilot) symbols transmitted at the beginning of each frame. The transmitter, before sending the data $\mathbf{x}$, sends a training sequence $\mathbf{x}_T = (x_{T,1}, \ldots, x_{T,N})$. According to the observation model (33), this sequence is affected by $h$, allowing the receiver to observe separately $\mathbf{y}_T = h\,\mathbf{x}_T + \mathbf{z}_T$, where $\mathbf{z}_T$ is the noise affecting the transmission of training symbols. The average energy of the training symbols is $P_T = \frac{1}{N}\mathbf{x}_T\mathbf{x}_T^\dagger$. Estimating $h$ in the ML sense given $\mathbf{y}_T$ and $\mathbf{x}_T$ amounts to minimizing $\|\mathbf{y}_T - h\mathbf{x}_T\|^2$ with respect to $h$. This yields $\hat{h} = \mathbf{y}_T\mathbf{x}_T^\dagger(\mathbf{x}_T\mathbf{x}_T^\dagger)^{-1} = h + \mathcal{E}$, where $\mathcal{E} = \mathbf{z}_T\mathbf{x}_T^\dagger(\mathbf{x}_T\mathbf{x}_T^\dagger)^{-1}$ denotes the estimation error. Next, we can write $\sigma_{\mathcal{E}}^2 = \mathrm{SNR}_T^{-1}$ with $\mathrm{SNR}_T \doteq \frac{NP_T}{\sigma_Z^2}$. Thus, the conditional pdf of $\hat{\boldsymbol{\theta}} = \hat{H}$ given $\boldsymbol{\theta} = H$ is the circularly complex Gaussian pdf $\mu_{\hat{\boldsymbol{H}}|\boldsymbol{H}} = C\mathcal{N}(H, \sigma_{\mathcal{E}}^2)$. Then, by using some algebra and the *a priori* distribution $\mu_{\boldsymbol{H}}$, the *a posteriori* distribution of $H$ given $\hat{H}$ can be expressed as

$$\mu_{\boldsymbol{H}|\hat{\boldsymbol{H}}}(H|\hat{H}) = \frac{\mu_{\hat{\boldsymbol{H}}|\boldsymbol{H}}(\hat{H}|H)\mu_{\boldsymbol{H}}(H)}{\int_{\Theta}\mu_{\hat{\boldsymbol{H}}|\boldsymbol{H}}(\hat{H}|H)d\mu_{\boldsymbol{H}}(H)} = C\mathcal{N}(\mu_{\hat{H}}, \delta\sigma_{\mathcal{E}}^2), \qquad (35)$$

where $\delta \doteq \frac{\mathrm{SNR}_T \sigma_H^2}{\mathrm{SNR}_T \sigma_H^2 + 1}$ and $\mu_{\hat{H}} \doteq \delta\hat{H} + (1-\delta)\mu_H$. An alternative expression of (35) in terms of the phase $\phi_H$ and the magnitude $r$ of $H = r\exp(j\phi_H)$ is given by

$$\mu_{r\phi_H|\hat{H}}(r, \phi_H|\hat{H} = \hat{h}) = \frac{r}{\pi\delta\sigma_{\mathcal{E}}^2}\exp\left(-\frac{r^2 - 2\delta|\mu_{\hat{H}}|r\cos(\phi_H - \phi_{\mu_{\hat{H}}}) + \delta^2|\mu_{\hat{H}}|^2}{\delta\sigma_{\mathcal{E}}^2}\right), \qquad (36)$$

where $\phi_{\mu_{\hat{H}}}$ denotes the phase of $\mu_{\hat{H}}$. The availability of the accuracy statistic (36) characterizing the channel estimation errors is the key feature to compute the EIO capacity.

## B. EIO Capacity of the Non-ergodic Ricean Fading Channel

Evaluating the EIO capacity (10) requires to solve an optimization problem where we have to determine the optimum set $\Lambda_{\mathrm{opt}} \subseteq \Theta$, and its associated channel state $h_{\mathrm{opt}} \in \Lambda_{\mathrm{opt}}$ minimizing the mutual information (11). However, in our case it can be observed that the mutual information computed with (34) only depends on the absolute value $|h|$ of the channel coefficient. Thus, the optimization over sets $\Lambda \subseteq \Theta$ of complex fading coefficients can be replaced with sets $\Lambda_I = \{h \in \Theta : |h| \in I\}$, where $I$ denotes an arbitrary positive real interval.

The conditional pdf $\mu_{r|\hat{H}}$ can be obtained by marginalizing (36), which results in the Ricean distribution

$$\mu_{r|\hat{H}}(r|\hat{H} = \hat{h}) = \frac{r}{\delta\sigma_{\mathcal{E}}^2/2}\exp\left(-\frac{r^2 + |\mu_{\hat{H}}|^2}{\delta\sigma_{\mathcal{E}}^2}\right)I_0\left(\frac{|\mu_{\hat{H}}|r}{\delta\sigma_{\mathcal{E}}^2/2}\right), \qquad (37)$$

where $I_0(\cdot)$ is the zero'th order *modified Bessel function of the first kind* [42, Eq. (8.445)]. Consequently, given an $\epsilon > 0$, the optimization problem now reduces to finding the optimum interval $I_{\text{opt}}^{(\epsilon)} \doteq [r_{\text{opt}}, 1/\epsilon]$ such that the set $\Lambda_{\text{opt}}^{(\epsilon)} \doteq \{h \in \Theta : |h| \in I_{\text{opt}}^{(\epsilon)}\}$ has probability $1 - \gamma$ (computed with (37)) when $\epsilon \to 0$. This follows from the fact that the mutual information is monotone increasing in $|h|$ while the intervals $I_{\text{opt}}^{(\epsilon)}$ are convex and compact, thus the infimum in the capacity expression actually equals the minimum over all $r$ ranging over the set $I_{\text{opt}}^{(\epsilon)}$. It follows that $r_{\text{opt}}^{(\epsilon)}(\gamma, \hat{h})$ is the $\gamma$-percentile given by $\Pr\left(H \in \Lambda_{\text{opt}}^{(\epsilon)} \mid \hat{H} = \hat{h}\right) = 1 - \gamma$. This probability can be computed from the pdf (37) as follows

$$\Pr\left(H \in \Lambda_{\text{opt}}^{(\epsilon)} | \hat{H} = \hat{h}\right) = \mathcal{Q}_1\left(\sqrt{\frac{2|\mu_{\hat{H}}|^2}{\delta \sigma_{\tilde{\mathcal{E}}}^2}}, \sqrt{\frac{2\left(r_{\text{opt}}^{(\epsilon)}(\gamma, \hat{h})\right)^2}{\delta \sigma_{\tilde{\mathcal{E}}}^2}}\right), \tag{38}$$

where $\mathcal{Q}_1(\alpha, \beta)$ is the first-order *Marcum Q-function* [43] (see Appendix IV). We note that the mutual information corresponding to the considered channel is maximized by using ZMCCG inputs with variance (transmit power) $P_{\hat{H}}$. Then, the EIO capacity can be shown to be given by

$$C_{EIO}(\gamma, \hat{h}, P_{\hat{H}}) = \log_2\left(1 + \frac{r_{\text{opt}}(\gamma, \hat{h})^2 P_{\hat{H}}}{\sigma_Z^2}\right), \tag{39}$$

by choosing $T = X$ in (10), where $r_{\text{opt}}^{(\epsilon)} \to r_{\text{opt}}$ with $\epsilon \to 0$.

We remark that for fixed $\delta > 0$, $\Pr\left(|H - \hat{H}| > \delta \mid \hat{H} = \hat{h}\right) \to 0$ as $N \to \infty$. Thus, any set $\Lambda_\delta = \{H \in \Theta : |H - \hat{h}| \le \delta\}$ contains a smaller and smaller neighborhood of the true parameter $H$ and hence by continuity $C_{EIO} \to \log_2\left(1 + \frac{|h|^2 P_{\hat{H}}}{\sigma_Z^2}\right)$ as $N \to \infty$. Therefore, we observe the expected result that the EIO capacity converges to the capacity with perfect CSI for all $\gamma \in [0, 1]$, as the training sequence length $N$ tends to infinity.

*C. Capacity of the Non-ergodic Ricean Fading Channel Based on Average Error Probability*

For comparison, we also evaluate the capacity expression (3) corresponding to the conventional notion of reliable communication explained in Section I, which is based on the average of the error probability over all channel estimation errors. We begin by computing the composite channel model (2), which contains the channel estimation errors

$$\begin{aligned} \mathbb{W}_{\hat{H}}(y|x) &= \mathbb{E}_{H|\hat{H}}\{W_H(y|x)|\hat{H}=\hat{h}\}, \\ &= \mathcal{CN}\big((\delta\hat{h} + (1-\delta)\mu_H)x, \sigma_Z^2 + \delta\sigma_{\mathcal{E}}^2|x|^2\big). \end{aligned} \qquad (40)$$

Then, it is not difficult to show that with Gaussian inputs the mutual information evaluated in this composite channel yields the following expression:

$$\begin{aligned} C(\hat{h}, P_{\hat{H}}) &= H(Y_{\hat{H}}|\hat{H}=\hat{h}) - H(Y_{\hat{H}}|X_{\hat{H}}, \hat{H}=\hat{h}), \\ &= \log_2\left(|\delta\hat{h} + (1-\delta)\mu_H|^2 P_{\hat{H}} + \sigma_Z^2 + \delta\sigma_{\mathcal{E}}^2 P_{\hat{H}}\right) - \mathbb{E}_{P_{\hat{H}}}\{\log_2(\sigma_Z^2 + \delta\sigma_{\mathcal{E}}^2|x|^2)\}, \\ &= \log_2\left(1 + \frac{|\delta\hat{h} + (1-\delta)\mu_H|^2 P_{\hat{H}}}{\sigma_Z^2 + \delta\sigma_{\mathcal{E}}^2 P_{\hat{H}}}\right) \\ &\quad + \left[\log_2\left(1 + \frac{\delta\sigma_{\mathcal{E}}^2 P_{\hat{H}}}{\sigma_Z^2}\right) - \exp\left(\frac{\sigma_Z^2}{\delta\sigma_{\mathcal{E}}^2 P_{\hat{H}}}\right) E_1\left(\frac{\sigma_Z^2}{\delta\sigma_{\mathcal{E}}^2 P_{\hat{H}}}\right)\right], \end{aligned} \qquad (41)$$

where the last equality follows by calculating the expectation and $E_1(\cdot)$ denotes the *exponential integral function* (see Appendix IV). Note that the first term in (41) provides an intuitive lower bound, i.e.,

$$C(\hat{h}, P_{\hat{H}}) \geq \log_2\left(1 + \frac{|\delta\hat{h} + (1-\delta)\mu_H|^2 P_{\hat{H}}}{\sigma_Z^2 + \delta\sigma_{\mathcal{E}}^2 P_{\hat{H}}}\right), \qquad (42)$$

which follows by upper bounding the second term in (41) using Jensen's inequality and the concavity of the log function. This lower bound parallels, for the considered estimation method, that found in (42).

*D. Achievable EIO Rates Associated to the Mismatched ML Decoder*

Mismatched decoding arises when the decoder is restricted to use a prescribed "metric" $d : \mathscr{X} \times \mathscr{Y} \to \mathbb{R}_+$, which does not necessarily match the true channel (cf. [44], [5], [10]). Given an output sequence $\mathbf{y} \in \mathscr{Y}^n$ and a channel estimate $\hat{h}$, we assume that decoding is performed by using the mismatched ML metric, i.e., we set $d_{\hat{H}}(\mathbf{x}_i, \mathbf{y}) = \|\mathbf{y} - \hat{h}\,\mathbf{x}_i\|^2$. Hence, the decoder declares that the codeword $\mathbf{x}_i \in \{\mathbf{x}_1, \ldots, \mathbf{x}_M\}$ was sent, iff $d_{\hat{H}}(\mathbf{x}_i, \mathbf{y}) < d_{\hat{H}}(\mathbf{x}_j, \mathbf{y})$ for all $j \neq i$, otherwise it declares an error. Obviously, suboptimal performance in terms of achievable EIO rates is expected for this decoder, since it does not necessarily achieve the EIO capacity. However, we aim at comparing the EIO capacity (39) (i.e. the ultimate limits) with the EIO rate $C_{\text{EIO-ML}}$ achievable with the mismatched ML decoder.

The expression of achievable EIO rates associated to the mismatched ML decoder can be obtained by combining the notion of EIO capacity with the previous results [5]

$$C_{\text{EIO-ML}}(\gamma, \hat{h}) = \max_{P_X \in \mathscr{P}_\Gamma} \sup_{\Lambda \subseteq \Theta:\, \Pr(\Lambda|\hat{H}=\hat{h}) \geq 1-\gamma} \inf_{\{H \in \Lambda,\, (\xi,\sigma) \in \mathcal{V}(H,\hat{h})\}} I(X_{\hat{H}}; Y_{\xi,\sigma} | \hat{H} = \hat{h}), \qquad (43)$$

where the set $\mathcal{V}(H, \hat{h}) \doteq \{(\xi, \sigma) \in (\Theta \times \mathbb{R}_+) : \mathbb{E}_{X_{\hat{H}} Y_{\xi,\sigma}}\{d_{\hat{H}}(x,y)\} \leq \mathbb{E}_{X_{\hat{H}} Y_H}\{d_{\hat{H}}(x,y)\}\}$, with $P_X V_{\xi,\sigma}(y) = P_X W_H(y)$ a.s.$\}$ and the mutual information is evaluated for an arbitrary channel $V_{\xi,\sigma}(\cdot|x) = \mathcal{CN}(\xi x, \sigma^2)$ with channel state $\xi \in \Theta$ and variance $\sigma^2$. Then, by computing the mutual information and the minimization set, it follows that

$$\begin{aligned}
C_{\text{EIO-ML}}(\gamma, \hat{h}) &= \inf_{\{H \in \Lambda,\, \mu \in \mathbb{C}:\, \text{Re}\{\xi \hat{h}\} \geq \text{Re}\{H\,\hat{h}\}\}} \log_2\left(1 + \frac{|\xi|^2 P_{\hat{H}}}{(|H|^2 - |\xi|^2) P_{\hat{H}} + \sigma_Z^2}\right), \\
&= \inf_{\{(r, \phi_H):\, H \in \Lambda\}} \log_2\left(1 + \frac{r^2 \cos^2(\phi_H - \phi_{\hat{H}}) P_{\hat{H}}}{r^2 \sin^2(\phi_H - \phi_{\hat{H}}) P_{\hat{H}} + \sigma_Z^2}\right), \qquad (44)
\end{aligned}$$

where the last equality follows by computing the minimizing value $\mu_{\text{opt}} = \dfrac{\text{Re}\{H\,\hat{h}\}}{|\hat{h}|^2}$ with the definitions of $H = r\exp(j\phi_H)$ and $\hat{h} = \hat{r}\exp(j\phi_{\hat{H}})$. It should be noted that, in contrast to the EIO capacity, here the achievable EIO rates associated to the mismatched ML decoder

are sensitive to phase errors between the channel and its estimate. For real-valued channels, mismatched ML decoding entails no performance loss since (44) equals the capacity (39), and thus the comparison with mismatched ML decoding would not make sense in that context.

Evaluating (43) further requires an evaluation of the optimal set $\Lambda_{\text{opt}} \subseteq \Theta$ of channel states maximizing (44) and whose probability is at least $1 - \gamma$. By inspecting expression (44), it is not difficult to see that this set is characterized by $\Lambda_{\text{opt}} = \{H \in \Theta : |\phi_H - \phi_{\hat{H}}| \leq \phi_{\mathcal{E}}, |H| \geq r_{\text{opt}}\}$ for some optimal values $(r_{\text{opt}}, \phi_{\mathcal{E}})$ guaranteeing that $\Pr(\Lambda_{\text{opt}}|\hat{H} = \hat{h}) = 1 - \gamma$ for a given channel estimate $\hat{h}$. We now need to evaluate the probability $\Pr(\Lambda_{\text{opt}}|\hat{H} = \hat{h})$, which consists in integrating the pdf (36) over the set $\Lambda_{\text{opt}}$ of complex values, resulting in the following integral expression

$$I(r_{\text{opt}}, \phi_{\mathcal{E}}) = \int_{r_{\text{opt}}}^{\infty} \int_{\phi_{\hat{H}} - \phi_{\mathcal{E}}}^{\phi_{\hat{H}} + \phi_{\mathcal{E}}} \mathbb{D}\, r \exp\left(-\mathbb{A}r^2 + \mathbb{B}r \cos(\phi_H - \phi_{\mu_{\hat{H}}})\right) dr d\phi_H, \qquad (45)$$

with constants $\mathbb{A} \doteq \dfrac{1}{\delta \sigma_{\mathcal{E}}^2}$, $\mathbb{B} \doteq \dfrac{2|\mu_{\hat{H}}|}{\sigma_{\mathcal{E}}^2}$ and $\mathbb{D} \doteq \dfrac{\mathbb{A}}{\pi} \exp\left(-\dfrac{\mathbb{B}^2}{4\mathbb{A}}\right)$. This integral can be numerically evaluated (see Appendix IV) and thus the rate expression (43) writes as

$$C_{\text{EIO-ML}}(\gamma, \hat{h}) = \min_{\{(r_{\text{opt}}, \phi_{\mathcal{E}}) : I(r_{\text{opt}}, \phi_{\mathcal{E}}) = 1 - \gamma\}} \log_2 \left(1 + \frac{r_{\text{opt}}^2 \cos^2(\phi_{\mathcal{E}}) P_{\hat{H}}}{r_{\text{opt}}^2 \sin^2(\phi_{\mathcal{E}}) P_{\hat{H}} + \sigma_Z^2}\right), \qquad (46)$$

where the minimization is taken over all pairs $(r_{\text{opt}}, \phi_{\mathcal{E}})$, defining the boundary of the region $\Lambda_{\text{opt}}$, for which $\Pr(\Lambda_{\text{opt}}|\hat{H} = \hat{h}) = 1 - \gamma$.

*E. Long-Term Power Allocation and Quantized CSI Feedback*

Next we concentrate on deriving optimal power allocation strategies $\{P_{\hat{H}} : \Theta \to \mathbb{R}_+\}$ for maximizing the mean EIO capacity under the long-term constraint $\mathbb{E}_{\hat{\mathbf{H}}}\{P_{\hat{H}}\} \leq \bar{P}$, for the cases of noiseless and noisy feedback. In this scenario, since each codeword experiences additive white Gaussian noise, random Gaussian codes with multiple codebooks are optimal. Based on the channel estimate available at the transmitter $\hat{\boldsymbol{\theta}} = \hat{H}$ (respectively its quantized value $\tilde{\boldsymbol{\theta}} = \tilde{H}$), a codeword is sent at a power level $P_{\hat{H}}$ (respectively $P_{\tilde{H}}$) given by the optimal power allocation function (cf. [9]). First consider the case of noiseless feedback (i.e. the transmitter knows $\hat{\boldsymbol{\theta}} = \hat{H}$). From (39) the mean EIO capacity writes as

$$\bar{C}_{EIO}(\gamma, \bar{P}) = \mathbb{E}_{\hat{\mathbf{H}}} \left\{ \sup_{P_{\hat{H}}: \mathbb{E}_{\hat{\mathbf{H}}}\{P_{\hat{H}}\} \leq \bar{P}} \log_2 \left( 1 + \frac{r_{\text{opt}}(\gamma, \hat{H})^2 P_{\hat{H}}}{\sigma_Z^2} \right) \right\}, \quad (47)$$

where the supremum is over all non-negative power allocation functions $P_{\hat{H}}$ such that $\mathbb{E}_{\hat{\mathbf{H}}}\{P_{\hat{H}}\} \leq \bar{P}$. Given a state measurement $\hat{H} = \hat{h}$, the transmitter selects a code with a power level $P_{\hat{H}}$ and uses $\hat{h}$ and $\mu_{r|\hat{H}}$ to compute (using (38)) the worst channel state $r^*(\gamma, \hat{h})$. Thus, the optimal power allocation maximizing (47) is easily derived as the well-known *water-filling* solution [9],

$$\frac{P_{\hat{H}}}{\sigma_Z^2} \doteq \left[ \frac{1}{r_0} - \frac{1}{r^*(\gamma, \hat{h})} \right]_+, \quad (48)$$

where $r_0$ is the positive constant guaranteeing the power constraint $\mathbb{E}_{\hat{\mathbf{H}}}\{P_{\hat{H}}\} = \bar{P}$ and $[x]_+ \doteq \max\{x, 0\}$.

Consider now the situation in which the receiver quantizes and sends to the transmitter the channel estimate $\hat{H}$, by using a rate-limited feedback link. Clearly, the performance is now a function of the amount of feedback bits $R_{\text{FB}}$. The receiver selects a quantized value among $M_{\text{FB}} = \lfloor 2^{2R_{\text{FB}}} \rfloor$ possibilities in the quantization codebook, which is assumed to be also known at the transmitter. This codebook is designed to minimize the MMSE between the input and its quantized value. We construct this codebook $Q[\hat{H}] \in \{\tilde{H}_1, \ldots, \tilde{H}_{M_{\text{FB}}}\}$ by using the non-uniform quantizer $Q[\,\cdot\,]$ designed with the well-known Lloyd-Max Algorithm [41]. We remark that the considered quantizer is not necessarily optimal for maximizing the EIO capacity. The reason is that the cost function (not necessary the MMSE) can exploit any channel invariance, which may be present in the communication model. Indeed, optimal design of quantized feedback is a vast topic and the literature is large and growing (see [45]–[49] and references therein). However, here we do not intend to design optimal feedback, the goal is to show how to incorporate limited feedback in the EIO capacity. Then, to capitalize on the rate-limited feedback the EIO capacity and its power allocation (48) should be modified accordingly.

Let $\tilde{H} = Q[\hat{H}]$ be the quantized value received at the transmitter corresponding to a channel estimate $\hat{H}$. In this case, the mean EIO capacity with rate-limited feedback is given by

$$\bar{C}_{EIO}(\gamma, \bar{P}) = \sup_{P_{\tilde{H}} : \mathbb{E}_{\tilde{\mathbf{H}}}\{P_{\tilde{H}}\} \leq \bar{P}} \sum_{i=1}^{M_{\text{FB}}} \log_2\left(1 + \frac{r_{\text{opt}}(\gamma, \tilde{h}_i)^2 P_{\tilde{H}_i}}{\sigma_Z^2}\right) \Pr(\tilde{H} = \tilde{h}_i), \quad (49)$$

where the supremum is over all non-negative power allocation functions $P_{\tilde{H}} : \{\tilde{H}_1, \ldots, \tilde{H}_{M_{\text{FB}}}\} \to \mathbb{R}_+$ such that $\sum_{i=1}^{M_{\text{FB}}} P_{\tilde{H}_i} \Pr(\tilde{H} = \tilde{h}_i) \leq \bar{P}$; here, $\Pr(\tilde{H} = \tilde{h}_i) \doteq \int_{\Lambda_{Q,i}} d\mu_{\hat{H}|\tilde{H}}(\hat{H}|\tilde{H} = \tilde{h}_i)$ where $\Lambda_{Q,i} \doteq \{\hat{h} \in \Theta : Q[\hat{h}] = \tilde{h}_i\}$ denotes the set of $\hat{h}$ yielding the quantized state $\tilde{h}_i$. The optimal value $r_{\text{opt}}(\gamma, \tilde{h}_i)$ in (49) can be computed by following the same steps as in (47) but according to the pdf $\mu_{r|\tilde{H}}$ given a quantized estimate $\tilde{h}$. It is immediate to see that the optimal power allocation must satisfy the power constraint with equality, and thus

$$\frac{P_{\tilde{H}_i}}{\sigma_Z^2} \doteq \left[\frac{1}{r_0} - \frac{1}{r_{\text{opt}}(\gamma, \tilde{h}_i)}\right]_+, \quad (50)$$

where $r_0$ is a positive constant ensuring the power constraint $\mathbb{E}_{\tilde{\mathbf{H}}}\{P_{\tilde{H}}\} = \bar{P}$.

It remains to compute the accuracy statistic represented by the pdf $\mu_{r|\tilde{H}}$ of $r = |h|$ given $\tilde{H} = \tilde{h}$, which characterizes the channel estimation and the quantization errors together, regarding the quantization method under consideration. In order to derive the accuracy statistic $\mu_{r|\tilde{H}}$, needed to compute the EIO capacity, we introduce the statistical model for quantization of channel estimates. From the rate distortion theory [50] and by considering the MMSE distortion, it is not difficult to see that $\mu_{\hat{H}|\tilde{H}=\tilde{h}} = \mathcal{CN}(\tilde{h}, \sigma_{\mathcal{E}_Q}^2)$, where the variance $\sigma_{\mathcal{E}_Q}^2$ corresponds to the quantization error of $\tilde{H}$, which is encoded with $R_{\text{FB}}$ bits per scalar symbol. According to this 32and expression (35), we can compute the pdf $\mu_{H|\tilde{H}} = \mathcal{CN}\big(\mu_{\tilde{H}}, \delta(\sigma_{\tilde{\mathcal{E}}}^2 + \delta\sigma_{\tilde{\mathcal{E}}_Q}^2)\big)$ with $\mu_{\tilde{H}} \doteq \delta\tilde{H} + (1-\delta)\mu_H$. Hence, the pdf required to derive the EIO capacity is given by

$$\mu_{r|\tilde{H}} = \frac{r}{\delta(\sigma_{\tilde{\mathcal{E}}}^2 + \delta\sigma_{\tilde{\mathcal{E}}_Q}^2)/2} \exp\left(-\frac{r^2 + |\mu_{\tilde{H}}|^2}{\delta(\sigma_{\tilde{\mathcal{E}}}^2 + \delta\sigma_{\tilde{\mathcal{E}}_Q}^2)}\right) I_0\left(\frac{|\mu_{\tilde{H}}|r}{\delta(\sigma_{\tilde{\mathcal{E}}}^2 + \delta\sigma_{\tilde{\mathcal{E}}_Q}^2)/2}\right), \qquad (51)$$

and its corresponding probability follows as

$$\Pr\big(H \in \Lambda_{\text{opt}} | \tilde{H} = \tilde{h}\big) = \mathcal{Q}_1\left(\sqrt{\frac{2|\mu_{\tilde{H}}|^2}{\delta(\sigma_{\tilde{\mathcal{E}}}^2 + \delta\sigma_{\tilde{\mathcal{E}}_Q}^2)}}, \sqrt{\frac{2r_{\text{opt}}^2(\gamma, \tilde{h})}{\delta(\sigma_{\tilde{\mathcal{E}}}^2 + \delta\sigma_{\tilde{\mathcal{E}}_Q}^2)}}\right). \qquad (52)$$

## V. NUMERICAL RESULTS

In this section, numerical results are presented based on the capacity expressions evaluated in Section IV, which correspond to different scenarios of a single-antenna *non-ergodic* Ricean fading channel.

We first assume communications without long-term power constraints (no power control is possible) and numerically evaluate: (i) the mean EIO with perfect feedback, i.e., $\hat{H}$ is available at both the transmitter and the receiver (expression (39)), (ii) the capacity corresponding to the conventional notion of reliability based on the average error probability (expression (41)), (iii) the EIO capacity without CSIT (expression (5)) and for comparison we also show the mean Shannon capacity with perfect CSI. Fig. 2(a) shows these quantities (in bits per channel use) versus the signal-to-noise ratio $\text{SNR} = \frac{|\mu_H|^2 \bar{P}}{\sigma_Z^2}$ for different outage probabilities $\gamma \in \{10^{-1}, 10^{-2}\}$. The Rice factor was $K_H = 0\,\text{dB}$, the power and the length of the training pilots are $P_T = \bar{P}$ and $N = 1$, respectively. We observe that the mean EIO capacity is quite large, in spite of the small training sequence. To achieve 2 bits with imperfect CSI ($\gamma = 0.01$) requires about $5.5\,\text{dB}$ more than in the case with perfect CSI. On the other hand, observe that the channel estimation errors are still quite large with a single pilot symbol (e.g. $\sigma_{\mathcal{E}}^2 = 1$ for SNR=$0\,\text{dB}$) and therefore the notion of reliability based on the average error probability yields much higher rates that may be not effectively supported in practical communication systems. It should be noted, however, that by choosing an outage probability $\gamma = 0.1$ and at an SNR of $15$ dB both notions of capacity lead to the same rate. This scenario has been outlined in the introduction section, exposing that the EIO capacity provides more precise control through $\gamma$ over the reliability function, at the expense of decreasing the information rate.

Fig. 2(b) compares the following capacities versus the SNR, for $\gamma = 10^{-2}$: (i) The mean EIO capacity with perfect feedback, (ii) the maximum EIO rate associated to the mismatched ML decoder (expression (46)), for different amounts of training and the mean Shannon capacity with perfect CSI. Observe that in order to achieve $2$ bits, a scheme using imperfect CSI and $N = 3$ pilot symbols (dot line) requires $7\,\text{dB}$, i.e., $4\,\text{dB}$ more than in the case with perfect CSI (solid line). If the number of pilot symbols is further reduced to $N = 1$, this gap increases to $5\,\text{dB}$. In comparison, mean EIO rates $\bar{C}_{\text{EIO-ML}}$ corresponding to the mismatched ML decoding are significantly smaller compared to the EIO capacity. Indeed, in order to achieve the same target rate, a communication system using the mismatched ML decoder would requires $2.5$ higher SNR. Thus, it follows that the accuracy of channel estimates provided by $N = \{1, 3\}$ pilot symbols is not enough to allow for reliable decoding with the mismatched ML decoder. However, if the number of pilot symbols is increase to $N = 10$ then this decoder can achieve rates close to the EIO capacity.

We now consider communications with long-term power constraints, so that power allocation functions are employed. The following scenarios are investigated: (i) the mean EIO capacity with perfect feedback and optimal power allocation (expression (47)), (ii) the mean EIO capacity with rate-limited feedback and optimal power allocation (expression (49)) and the ergodic capacity with perfect CSI. Fig. 2(c) shows the mean EIO capacity for $\gamma = 0.01$ and rate-limited feedback/CSIT versus the SNR. It is seen that the mean EIO rates increase with the amount of feedback bits. In the case of the ergodic capacity (perfect CSI with power allocation) the SNR requirement for 2 bits is $2\,\text{dB}$ (solid line), while $3\,\text{dB}$ are required for the mean Shannon capacity (no power allocation), $7\,\text{dB}$ for the mean EIO (imperfect CSI and power allocation) with perfect feedback ($N = 1$ and $\gamma = 0.01$), and $9.5\,\text{dB}$ without power allocation. Thus, in presence of a long-term power constraint the gap between the EIO capacity and the ergodic capacity is slightly smaller than without such constraint, i.e., $5\,\text{dB}$ using power allocation as opposed to $6.5\,\text{dB}$ without it. We observe that with rate-limited feedback larger gains can be obtained by increasing the SNR. The gap between the mean EIO capacity for $1$ bit of rate-limited feedback and that for to $3$ bits is 5dB (at a capacity of 2 bits), and this gap increases with the SNR. In particular, a scheme using $R_{FB} = 3$ bits of feedback achieves almost the same performance as perfect feedback. Therefore, using this information a system designer may decide the number of feedback bits required to achieve certain target rates.

Finally, we study the impact of the imperfect CSI on the mean EIO capacity for different fading statistics, i.e., different Rice factors, and perfect feedback. Fig. 2(d) shows the mean EIO capacity for Rice factors $K_H \in \{-15, 0, 25\}\,\text{dB}$ and amounts of training $N \in \{1, 3\}$. We observe that increasing the Rice factor increases the impact of the estimation errors on the mean EIO capacity. For high value of $K_H = 25\,\text{dB}$ (i.e. smaller variance $\sigma_H^2$) the mean EIO capacity is not sensitive to the amount of training. In contrast, for the smaller Rice factor $K_H = -15\,\text{dB}$ it is more important to achieve accurate channel estimates. This observation can be understood from the notion of EIO capacity that depends on the trade-off between the estimation error $\sigma_{\tilde{\varepsilon}}^2$ and the variance of the fading process $\sigma_H^2$ (see expression (35)). Such analysis could serve as a basis to decide in practical situations whether or not, depending on the nature of the fading process, robust channel estimation is needed. The worst case is observed for the range of intermediate Rice factors (i.e. $K_H = 0 \, \text{dB}$) since for these values the uncertainty about the accuracy of estimates is maximal.

## VI. Summary and Discussion

In this paper we investigated the problem of reliable transmission over discrete memoryless channels (DMCs) when the receiver and the transmitter only know noisy estimates of the time-varying states and fixed states controlling the communication. We proposed to characterize the information theoretical limits of such scenarios in terms of the novel notion of *estimation-induced outage (EIO) capacity*. In this setting, the goal of the transmitter and the receiver is to construct codes, based on accuracy statistics for the channel states, to guarantee the desired communication service (achieving target rates with small error probability) whatever the quality of the estimates during the transmission. We provided a single-letter characterization of the optimal trade-off between the maximum achievable EIO rate and the outage probability (the QoS), by proving an associated coding theorem and its strong converse. The EIO capacity can be viewed as unification of several useful capacity notions for memoryless channel models with uncertainty regarding the channel states.

A *non-ergodic* Ricean fading model is used to illustrate the above results by computing its mean EIO capacity. These results are useful for a system designer to assess the amount of training and feedback required to achieve target rates over a DMC with a given channel statistic. The maximum achievable EIO rate with Gaussian codebooks of a naive system whose receiver uses the mismatched ML decoder based on the channel estimate was also studied. Our results indicate that if the channel estimates are not precise enough (e.g. the training phase is too short) then this decoder can be largely suboptimal for the considered scenario. An improved decoder should use a metric based on maximum *a posteriori* (MAP) probability [19], [51], i.e. taking into account the statistical nature of the state estimation errors. Moreover, the study of practical coding schemes satisfying the outage constraints, which perform close to the theoretical optimum given by the EIO capacity, is also a topic of interest.

Possible direct applications of these results arise in practical communication systems with small training overhead and QoS constraints, such as OFDM or some MIMO systems. Another application scenario arises in the context of cellular coverage, where the average of EIO capacity would characterize performance over multiple communication sessions of many users with different geographic locations [52]. In that scenario, the system designer must guarantee a QoS during the connection session, i.e., reliable communication for $(1 - \gamma)$-percent of users, even for users with poor channel estimates. As a more challenging problem, it would be interesting to extend the EIO capacity to multiuser channels (e.g. MIMO broadcast channel and MIMO multiple access channel) with imperfect CSI.

## ACKNOWLEDGMENT

The authors would like to thank Prof. Te Sun Han for helpful discussions. They are also grateful to the Associate Editor Prof. Holger Boche and the anonymous reviewers, whose constructive comments greatly improved the presentation of the paper.

## APPENDIX I

## INFORMATION-TYPICAL SETS AND BASIC PROPERTIES

Information (or Kullback-Leibler) divergence of PDs can be interpreted as a (non-symmetric) analogue of Euclidean distance [53]. It entails the definition of *I-typical* sets, first suggested by

37Csiszár and Narayan [54]. Several results for standard "strongly typical sets" can be extended to "information-typical sets" [35].

Throughout the appendices we use the following notation. The empirical PD $\hat{P}_n(\mathbf{x};\cdot) \in \mathscr{P}(\mathscr{X})$ associated with a sample sequence $\mathbf{x} = (x_1,\ldots,x_n) \in \mathscr{X}^n$ is $\hat{P}_n(\mathbf{x};\mathscr{A}) = N(\mathscr{A}|\mathbf{x})/n$ with $N(\mathscr{A}|\mathbf{x}) = \sum_{i=1}^{n} 1_{\mathscr{A}}(x_i)$, and $\widehat{W}_n(\mathbf{x},\mathbf{y};\cdot|a)$ is the empirical conditional PD associated with $\mathbf{x}$ and $\mathbf{y} = (y_1,\ldots,y_n) \in \mathscr{Y}^n$, for $a \in \mathscr{X}$. The set $\mathscr{P}_n(\mathscr{X}) \subset \mathscr{P}(\mathscr{X})$ denotes the set of all rational point probability masses on $\mathscr{X}$, and its cardinality is bounded by $\|\mathscr{P}_n(\mathscr{X})\| \leq (1+n)^{\|\mathscr{X}\|}$ [40]. We shall use the total variation or variational distance defined by $\mathcal{V}(P,Q) = 2\sup_{\mathscr{A} \subseteq \mathscr{X}}|P(\mathscr{A}) - Q(\mathscr{A})|$. Pinsker's inequality [40] for conditional PDs states that $\mathcal{V}(W \circ P, V \circ P) \leq \sqrt{\mathcal{D}(W\|V|P)/2}$. Let $Q, P_X \in \mathscr{P}(\mathscr{X})$, then $Q$ is said to be absolutely continuous with respect to $P_X$, denoted $Q \ll P_X$, if $Q(\mathscr{A}) = 0$ for every set $\mathscr{A} \subset \mathscr{X}$ for which $P_X(\mathscr{A}) = 0$. The *support* of a conditional PD $\{W: \mathscr{X} \longmapsto \mathscr{Y}\} \in \mathscr{P}(\mathscr{Y})$ with respect to the PD $P_X$ is defined as the set $S_P(W) = \{y \in \mathscr{Y}: W(y|x) > 0 \text{ for all } P_X(x) > 0\}$. Let $\{W, V: \mathscr{X} \longmapsto \mathscr{Y}\} \in \mathscr{P}(\mathscr{Y})$ be two conditional PDs, then $\{W\}$ is said to be absolutely continuous with respect to $\{V\}$, writes $W \ll V$, if $S_P(W) \subseteq S_P(V)$. Thus, it follows that $\mathcal{D}(W\|V|P_X) < \infty$ iff $W \ll V$. Let $\mathcal{W}_\Lambda$ be a convex set of PDs $\{W_\theta: \mathscr{X} \longmapsto \mathscr{Y}\}_{\theta \in \Lambda} \in \mathscr{P}(\mathscr{Y})$; there is one PD whose support contains all the others' supports which is called the *support* of the set $\mathcal{W}_\Lambda$ and is denoted by $S_P(\mathcal{W}_\Lambda)$.

*Definition 1.1 (Set of types):* For any PD $P_n \in \mathscr{P}_n(\mathscr{X})$, the set of sequences $\mathbf{x} \in \mathscr{X}^n$ with type $P_n$ is defined by $\mathcal{T}^n_{[P_n]} \doteq \{\mathbf{x} \in \mathscr{X}^n: \mathcal{D}(\hat{P}_n\|P_n) = 0\}$, where $\hat{P}_n(\mathbf{x},\cdot)$ is the empirical PD. Similarly, for a conditional PD $\widehat{W}_n(\cdot|x) \in \mathscr{P}_n(\mathscr{Y})$, the set of sequences $\mathbf{y} \in \mathscr{Y}^n$ with type $W_n$ is defined by $\mathcal{T}^n_{[W_n]}(\mathbf{x}) \doteq \{\mathbf{y} \in \mathscr{Y}^n: \mathcal{D}(\widehat{W}_n\|W_n|\hat{P}_n) = 0\}$ for each $\mathbf{x} \in \mathscr{X}^n$ and $\widehat{W}_n(\mathbf{x},\mathbf{y};b|a)N(a|\mathbf{x}) = N(a,b|\mathbf{x},\mathbf{y})$ is the empirical conditional PD.

*Definition 1.2 (Set of I-typical sequences):* For any PD $P_X \in \mathscr{P}(\mathscr{X})$, the set of sequences

June 9, 2018             To appear in IEEE Transactions on Information Theory

$\mathbf{x} \in \mathscr{X}^n$, called *I-typical* with constant $\delta > 0$, is defined by $\mathcal{T}^n_{[X]_\delta} \doteq \{\mathbf{x} \in \mathscr{X}^n : \mathcal{D}(\hat{P}_n \| P_X) \leq \delta\}$, where $\hat{P}_n$ is the empirical PD such that $\hat{P}_n \ll P_X$. Similarly, for a conditional PD $\{W : \mathscr{X} \longmapsto \mathscr{Y}\} \in \mathscr{P}(\mathscr{Y})$, the set of sequences $\mathbf{y} \in \mathscr{Y}^n$ conditioned to $\mathbf{x} \in \mathscr{X}^n$, called *conditional I-typical* with constant $\delta > 0$, is defined by $\mathcal{T}^n_{[Y|X]_\delta}(\mathbf{x}) \doteq \{\mathbf{y} \in \mathscr{Y}^n : \mathcal{D}(\widehat{W}_n \| W | \hat{P}_n) \leq \delta\}$, where $\widehat{W}_n$ is the empirical conditional PD such that $\widehat{W}_n \ll W$ (respect to $\hat{P}_n$).

*Lemma 1.3 (Uniform continuity of the entropy function):* Let $P, Q \in \mathscr{P}(\mathscr{X})$ be two PDs and $\{W, V : \mathscr{X} \longmapsto \mathscr{Y}\} \in \mathscr{P}(\mathscr{Y})$ be two conditional PDs. Then, from Lemma 1.2.7 [40],

(i) If $\mathcal{V}(P, Q) \leq \Theta \leq 1/2$, $\Rightarrow$ $|H(X_P) - H(X_Q)| \leq -\Theta \log \frac{\Theta}{\|\mathscr{X}\|}$.

(ii) If $\mathcal{V}(V \circ P, W \circ P) \leq \Theta \leq 1/2$, $\Rightarrow$ $|H(Y_V|X_P) - H(Y_W|X_P)| \leq -\Theta \log \frac{\Theta}{\|\mathscr{X}\|\|\mathscr{Y}\|}$.

*Proposition 1.4: (Properties of I-typical sequences)*

(i) Any sequence $\mathbf{x} \in \mathcal{T}^n_{[X]_\delta}$ implies $\mathcal{V}(\hat{P}_n, P) \leq \sqrt{\delta/2}$. Moreover any sequence $\mathbf{y} \in \mathcal{T}^n_{[Y|X]_\delta}(\mathbf{x})$ implies $\mathcal{V}(\widehat{W}_n \circ \hat{P}_n, W \circ \hat{P}_n) \leq \sqrt{\delta/2}$, for all $\mathbf{x} \in \mathscr{X}^n$.

(ii) There exists sequences $(\delta_n)_{n \in \mathbb{N}_+}$ and $(\delta'_n)_{n \in \mathbb{N}_+}$ in $\mathbb{R}_+$ (which only depend on $\|\mathscr{X}\|, \|\mathscr{Y}\|$) such that $(\delta_n, \delta'_n) \to 0$ and $n \log^{-1}(n+1)(\delta_n, \delta'_n) \to \infty$ as $n \to \infty$, so that for every $P_X \in \mathscr{P}(\mathscr{X})$ and $\{W : \mathscr{X} \longmapsto \mathscr{Y}\} \in \mathscr{P}(\mathscr{Y})$, $P^n_X(\mathcal{T}^n_{[X]_{\delta_n}}) > 1 - \epsilon_n$ and $W^n(\mathcal{T}^n_{[Y|X]_{\delta'_n}} | \mathbf{x}) > 1 - \epsilon'_n$ with

$$\epsilon_n = \exp\{-n(\delta_n - n^{-1}\|\mathscr{X}\|\log(n+1))\},$$

$$\epsilon'_n = \exp\{-n(\delta'_n - n^{-1}\|\mathscr{X}\|\|\mathscr{Y}\|\log(n+1))\}.$$

Note that $\log(n+1) < \sqrt{n}$ and consequently these sequences converge to zero with a rate higher than that obtained for strongly typical sets [35].

(iii) Given $P, Q \in \mathscr{P}(\mathscr{X})$ and $\{W, V : \mathscr{X} \longmapsto \mathscr{Y}\} \in \mathscr{P}(\mathscr{Y})$ and $\delta > 0$. Then, we have that

$$\text{if} \quad \mathcal{D}(Q\|P) \leq \delta \quad \Rightarrow \quad |H(X_Q) - H(X_P)| \leq -\sqrt{\delta/2} \log \frac{\sqrt{\delta/2}}{\|\mathscr{X}\|},$$

$$\text{if} \quad \mathcal{D}(W\|V|P) \leq \delta \quad \Rightarrow \quad |H(Y_W|X_P) - H(Y_V|X_P)| \leq -\sqrt{\delta/2} \log \frac{\sqrt{\delta/2}}{\|\mathscr{X}\|/|\mathscr{Y}\|}.$$

(iv) There exists sequences $(\epsilon_n)_{n \in \mathbb{N}_+}$ and $(\epsilon'_n)_{n \in \mathbb{N}_+}$ in $\mathbb{R}_+$ with $(\epsilon_n, \epsilon'_n) \to 0$, as well as in (ii), so that for every $P_X \in \mathscr{P}(\mathscr{X})$ and $\{W : \mathscr{X} \longmapsto \mathscr{Y}\} \in \mathscr{P}(\mathscr{Y})$, we have that

$$\left|\frac{1}{n} \log \|\mathcal{T}^n_{[X]_{\delta_n}}\| - H(X)\right| \leq \epsilon_n,$$

$$\left|\frac{1}{n} \log \|\mathcal{T}^n_{[Y|X]_{\delta'_n}}(\mathbf{x})\| - H(Y|X)\right| \leq \epsilon'_n, \quad \text{for each } \mathbf{x} \in \mathcal{T}^n_{[X]_{\delta_n}}.$$

*Proof:* Assertion (i) immediately follows from Pinsker's inequality. Assertion (iii) follows from (i) and Lemma 1.3. Assertion (iv) immediately follows by defining I-typical sets using $(\delta_n, \delta'_n)$-sequences and from the claim (iii), where the existence of such sequences was proved in (ii). In order to prove the claim (ii), it is sufficient to show the second expression:

$$\begin{aligned}
W^n\big(\mathscr{Y}^n \setminus \mathcal{T}^n_{[Y|X]_{\delta'_n}}|\mathbf{x}\big) &= \sum_{\{V_n : \mathcal{D}(V_n\|W|\hat{P}_n) > \delta'_n, V_n \ll W\}} W^n\big(\mathcal{T}^n_{[V_n]}|\mathbf{x}\big) \\
&\leq \sum_{\{V_n : \mathcal{D}(V_n\|W|\hat{P}_n) > \delta'_n, V_n \ll W\}} \exp\big[-n\mathcal{D}(V_n\|W|\hat{P}_n)\big] \\
&\leq (1+n)^{\|\mathscr{X}\|\|\mathscr{Y}\|} \exp(-n\delta'_n) \doteq \epsilon',
\end{aligned}$$

for each $\mathbf{x} \in \mathscr{X}^n$. ∎

*Lemma 1.5 (Uniform continuity of I-divergences):* (i) Given conditional PDs $\{W, Z, V : \mathscr{X} \longmapsto \mathscr{Y}\} \in \mathscr{P}(\mathscr{Y})$ such that $W, Z \ll V$ with respect to some PD $P_X \in \mathscr{P}(\mathscr{X})$. For each $\epsilon > 0$, if $\mathcal{D}(Z\|W|P_X) \leq \epsilon$ there exists $\delta > 0$ such that $|\mathcal{D}(Z\|V|P_X) - \mathcal{D}(W\|V|P_X)| \leq \delta$ with $\delta \doteq -\sqrt{\epsilon/2} \log\big(\sqrt{\epsilon/2}/(\|\mathscr{X}\|\|\mathscr{Y}\|^2)\big) \to 0$ as $\epsilon \to 0$.

(ii) Similarly, given PDs $Q, Z, P_X \in \mathscr{P}(\mathscr{X})$ such that $Q, Z \ll P_X$. For each $\epsilon > 0$, if $\mathcal{D}(Z\|Q) \leq \epsilon$ then there exists $\delta' > 0$ such that $|\mathcal{D}(Z\|P_X) - \mathcal{D}(Q\|P_X)| \leq \delta'$ with $\delta' \doteq -\sqrt{\epsilon/2} \log\big(\sqrt{\epsilon/2}/\|\mathscr{X}\|^2\big) \to 0$ as $\epsilon \to 0$.

*Proof:* We only prove the first statement, since then (ii) follows immediately from (i). We observe that, from Proposition 1.4 (i) and Lemma 1.3, $\mathcal{D}(Z\|W|P_X) \leq \epsilon$ implies $|H(Y_Z|X_P) - H(Y_W|X_P)| \leq -\sqrt{\epsilon/2}\log\frac{\sqrt{\epsilon/2}}{\|\mathcal{X}\|\|\mathcal{Y}\|}$. Thus, the poof follows by considering the inequalities

$$\begin{aligned}|\mathcal{D}(Z\|V|P_X) - \mathcal{D}(W\|V|P_X)| &\leq |H(Y_Z|X_P) - H(Y_W|X_P)| \\ &+ \sum_{a\in\mathcal{X}}\sum_{b\in\mathcal{Y}} P_X(a)|Z(b|a) - W(b|a)|\log\|\mathcal{Y}\| \\ &\leq -\sqrt{\epsilon/2}\log\left(\sqrt{\epsilon/2}/(\|\mathcal{X}\|\|\mathcal{Y}\|)\right) + \sqrt{\epsilon/2}\log\|\mathcal{Y}\| \doteq \delta.\end{aligned}$$

∎

*Lemma 1.6 (Large probability of I-typical sets):* Let $\mathcal{T}^n_{[X]_{\delta_n}}$ and $\mathcal{T}^n_{[Y|X]_{\delta_n}}(\mathbf{x})$ be I-typical and conditional I-typical sets, respectively. The probability that a sequence does not belong to these sets converge to zero,

$$\lim_{n\to\infty} P^n\Big(\mathcal{X}^n \setminus \mathcal{T}^n_{[X]_{\delta_n}}\Big) = 0, \quad \lim_{n\to\infty} W^n\Big(\mathcal{Y}^n \setminus \mathcal{T}^n_{[Y|X]_{\delta_n}}|\mathbf{x}\Big) = 0, \quad \text{for every } \mathbf{x} \in \mathcal{X}^n.$$

Furthermore, $\mathcal{D}(\hat{P}_n\|P_X) \to 0$ and $\mathcal{D}(\widehat{W}_n\|W|\hat{P}_n) \to 0$ as $n \to \infty$ with probability 1.

*Proof:* We observe from assertion (ii) in Proposition 1.4 that

$$W^n\big([\mathcal{T}^n_{[Y|X]_{\delta_n}}]^c|\mathbf{x}\big) \leq \exp\big[-n\big(\delta_n - n^{-1}\|\mathcal{X}\|\|\mathcal{Y}\|\log(n+1)\big)\big],$$

for every $\mathbf{x} \in \mathcal{X}^n$, and then it expression goes to zero as $n \to \infty$. The second assertion follows from the fact that $\sum_{n=1}^{\infty} W^n\big(\{\mathbf{y} \in \mathcal{Y}^n : \mathcal{D}(\widehat{W}_n\|W|\hat{P}_n) > \delta_n\}|\mathbf{x}\big) < \infty$, and by applying the Borel-Cantelli Lemma [55], we obtain $\Pr\Big(\limsup_{n\to\infty}\{\mathcal{D}(\widehat{W}_n\|W|\hat{P}_n) > \delta_n\}\big|\mathbf{x}\Big) = 0$, which concludes the proof. ∎

*Lemma 1.7:* Given $0 < \eta < 1$, $P_X \in \mathcal{P}(\mathcal{X})$ and the set of conditional PDs $\{W_\theta : \mathcal{X} \longmapsto \mathcal{Y}\}_{\theta\in\Lambda}$ for some set $\Lambda \subset \Theta$. Then there exist sequences $(\epsilon_n)_{n\in\mathbb{N}_+}$ and $(\epsilon'_n)_{n\in\mathbb{N}_+}$ in $\mathbb{R}_+$ with $(\epsilon_n, \epsilon'_n) \to 0$, which only depend on $\|\mathcal{X}\|$, $\|\mathcal{Y}\|$ and $\eta$, so that:

(i) if $\inf_{\theta \in \Lambda} W_\theta P_X^n(\mathscr{A}) \geq \eta$ for $\mathscr{A} \subset \mathscr{X}^n$, then $\frac{1}{n} \log \|\mathscr{A}\| \geq \sup_{\theta \in \Lambda} H(Y_\theta) - \epsilon_n$,

(ii) if $\inf_{\theta \in \Lambda} W_\theta^n(\mathscr{B}|\mathbf{x}) \geq \eta$ for $\mathscr{B} \subset \mathscr{Y}^n$ and $\mathbf{x} \in \mathcal{T}_{[X]_\delta}^n$, then $\frac{1}{n} \log \|\mathscr{B}\| \geq \sup_{\theta \in \Lambda} H(Y_\theta|X) - \epsilon_n'$.

This Lemma simply follows from the proof of Corollary 1.2.14 in [40] and the previous Lemmata.

## APPENDIX II

### AUXILIARY RESULTS

This appendix introduces a few concepts shedding more light on the encoder and decoder required to achieve the EIO capacity and furthermore provides some auxiliary technical results needed for the proof of the Generalized Maximal Code Lemma 3.2 in Section III.

*Unfeasibility of Mismatched Typical Decoding:* Consider a DMC and its (noisy) estimate $\{\mathbb{W}_\theta, \mathbb{W}_{\hat{\theta}} : \mathscr{T} \longmapsto \mathscr{Y} \times \mathscr{V}\}$. The following Lemma shows that an I-typical decoder based on $\{\mathbb{W}_{\hat{\theta}}\}$ yields an error probability that approaches one when $\mathbb{W}_{\hat{\theta}} \neq \mathbb{W}_\theta$. This reveals that conventional I-typical sets with respect to $\{\mathbb{W}_{\hat{\theta}}\}$ merely specify some local structure in a small neighborhood of $\{\mathbb{W}_\theta\}$ but not in the whole space (as outlined in [56]). This fact does not establish that any I-typical decoder is not useful for decoding with imperfect CSI, but it shows that there are no decoding sets $\{\mathscr{D}_i^n \subseteq \mathcal{T}_{[Y_{\hat{\theta}} V_{\hat{\theta}}|T]_{\delta_n}}^n(\mathbf{t}_i)\}$ and codewords $\{\mathbf{t}_i\} \subseteq \mathcal{T}_{[T]_{\delta_n}}^n$ such that $\mathbb{W}_\theta^n(\mathscr{D}_i^n|\mathbf{t}_i) > 1 - \epsilon$ for all $n \geq n_0$.

*Lemma 2.1:* Let $\{\mathbb{W}, \mathbb{V} : \mathscr{T} \longmapsto \mathscr{Y} \times \mathscr{V}\}$ be two channels such that $\mathcal{D}(\mathbb{W}\|\mathbb{V}|P_T) > \xi > 0$ and $\mathbb{W} \ll \mathbb{V}$ respect to an arbitrary PD $P_T \in \mathscr{P}(\mathscr{T})$. Let $\mathcal{T}_{[Y_\mathbb{W}|T]_{\delta_n}}^n(\mathbf{t})$ and $\mathcal{T}_{[Y_\mathbb{V}|T]_{\delta_n}}^n(\mathbf{t})$ denote the corresponding associated conditional I-typical sets, for every $\mathbf{t} \in \mathcal{T}_{[T]_{\delta_n}}^n$. Then, there exists an index $n_0 \in \mathbb{N}_+$ such that for all $n \geq n_0$ these sets are disjoint and thus $\mathbb{W}^n\big(\mathcal{T}_{[Y_\mathbb{V}|T]_{\delta_n}}^n\big|\mathbf{t}\big) \to 0$ as $n \to \infty$. Furthermore, the quantity $\mathcal{D}(\widehat{W}_n\|\mathbb{V}|\hat{P}_n) \to \mathcal{D}(\mathbb{W}\|\mathbb{V}|P_T)$ with $\mathbb{W}^n$-probability 1 as $n \to \infty$.

*Proof:* We must show that given arbitrary $\mathbb{W}, \mathbb{V}, P_T$ such that $\mathcal{D}(\mathbb{W}\|\mathbb{V}|P_T) > \xi > 0$ with $\mathbb{W} \ll \mathbb{V}$, then for every sequences $(\mathbf{y}, \mathbf{v}) \in \mathcal{T}^n_{[Y_\mathbb{W} V_\mathbb{W}|T]_{\delta_n}}(\mathbf{t})$ (i.e. $\mathcal{D}(\widehat{W}_n\|\mathbb{W}|\hat{P}_n) \leq \delta_n$ with $\widehat{W}_n \ll \mathbb{W}$) and each $\mathbf{t} \in \mathcal{T}$, there exists $n_0 = n_0(\|\mathcal{T}\|, \|\mathcal{Y}\|, \|\mathcal{V}\|, \delta_n, \xi) \in \mathbb{N}_+$ such that $\mathcal{D}(\widehat{W}_n\|\mathbb{V}|\hat{P}_n) > \delta_n$ for all $n \geq n_0$, which implies that $(\mathbf{y}, \mathbf{v}) \in \mathcal{T}^n_{[Y_\mathbb{W} V_\mathbb{W}|T]_{\delta_n}}(\mathbf{t}) \cap [\mathcal{T}^n_{[Y_\mathbb{V} V_\mathbb{V}|T]_{\delta_n}}(\mathbf{t})]^c$. To this end, we know from Lemma 1.5 that $\mathcal{D}(\widehat{W}_n\|\mathbb{W}|\hat{P}_n) \leq \delta_n$ implies $|\mathcal{D}(\widehat{W}_n\|\mathbb{V}|\hat{P}_n) - \mathcal{D}(\mathbb{W}\|\mathbb{V}|P_T)| \leq \delta'_n$, with $\delta'_n \doteq -\sqrt{\delta_n/2}\log\left(\sqrt{\delta_n/2}/(\|\mathcal{T}\|\|\mathcal{Y}\|^3\|\mathcal{V}\|^3)\right) < \xi$ provided for all $n \geq n_0$. We have also used the fact that $\mathbf{t} \in \mathcal{T}^n_{[T]_{\delta_n}}$ and thus $\mathcal{D}(\hat{P}_n\|P_T) \leq \delta_n$, yielding to $|\mathcal{D}(\mathbb{W}\|\mathbb{V}|\hat{P}_n) - \mathcal{D}(\mathbb{W}\|\mathbb{V}|P_T)| \leq \sqrt{2\delta_n}\log\|\mathcal{Y}\|\|\mathcal{V}\|$ for all $n \geq n_0$.

Hence, it follows that $\mathcal{D}(\widehat{W}_n\|\mathbb{V}|\hat{P}_n) \geq \mathcal{D}(\mathbb{W}\|\mathbb{V}|P_T) - \delta'_n > \xi - \delta'_n$. For instance, for any $\xi > 0$ there exits $n_0 \in \mathbb{N}_+$ such that for all $n \geq n_0$ $\mathcal{D}(\widehat{W}_n\|\mathbb{V}|\hat{P}_n) > \delta_n \doteq \xi - \delta'_n$ ($\to 0$ as $n \to \infty$), which implies that $(\mathbf{y}, \mathbf{v}) \in \mathcal{T}^n_{[Y_\mathbb{W} V_\mathbb{W}|T]_{\delta_n}}(\mathbf{t}) \cap [\mathcal{T}^n_{[Y_\mathbb{V} V_\mathbb{V}|T]_{\delta_n}}(\mathbf{t})]^c$. Finally, since $\mathcal{T}^n_{[Y_\mathbb{V} V_\mathbb{V}|T]_{\delta_n}}(\mathbf{t}) \subseteq [\mathcal{T}^n_{[Y_\mathbb{W} V_\mathbb{W}|T]_{\delta_n}}(\mathbf{t})]^c$ for every $\mathbf{t} \in \mathcal{T}^n_{[T]_{\delta_n}}$, we have from Lemma 1.6 that $\mathbb{W}^n\left(\mathcal{T}^n_{[Y_\mathbb{V} V_\mathbb{V}|T]_{\delta_n}}|\mathbf{t}\right) \to 0$ as $n \to \infty$, concluding the proof of the first claim. To prove the second assertion, from Lemma 1.5 and the last assertion we can see that for every $\eta > 0$ there exits $n_0 \in \mathbb{N}_+$ and $(\delta_n)_{n \in \mathbb{N}_+}$ such that the set $\mathscr{B}^n_\eta(\mathbf{t}) \doteq \left\{(\mathbf{y}, \mathbf{v}) \in \mathcal{Y}^n \times \mathcal{V}^n : |\mathcal{D}(\widehat{W}_n\|\mathbb{V}|\hat{P}_n) - \mathcal{D}(\mathbb{W}\|\mathbb{V}|P_T)| > \eta, \widehat{W}_n \ll \mathbb{V}\right\} \subseteq [\mathcal{T}^n_{[Y_\mathbb{W} V_\mathbb{W}|T]_{\delta_n}}(\mathbf{t})]^c$. Hence, $\Pr\left(\mathscr{B}^n_\eta|\mathbf{t}\right) \leq \epsilon_n$ with $\epsilon_n(\delta_n) \to 0$ as $n \to \infty$, which means that $\sum_{n=n_0}^{\infty} \Pr\left(\mathscr{B}^n_\eta(\mathbf{t})|\mathbf{t}\right)$ converges for all $\eta > 0$. The proof concludes by the Borel-Cantelli Lemma [55]. ∎

We now construct a formal definition of the decoding sets used to achieve the EIO capacity.

*Definition 2.2 (Robust $\epsilon$-decoding sets):* Let $\mathcal{T}_0 \subset \mathcal{T}^n$ denote a set of transmit sequences. A set $\mathscr{B} \subset \mathcal{Y}^n \times \mathcal{V}^n$ is called a *robust $\epsilon$-decoding set* for a sequence $\mathbf{t} \in \mathcal{T}_0$ and an unknown DMC $\left\{\mathbb{W}_\theta : \mathcal{T} \times \Theta \longmapsto \mathcal{Y} \times \mathcal{V}\right\}$ with $\theta \in \Theta$ if the conditional (w.r.t. $\hat{\theta}$) probability of $\theta$, for which the $\mathbb{W}^n_\theta$-probability of $\mathscr{B}$ exceeds $1-\epsilon$, is at least $1-\gamma$, i.e., $\Pr\left(\{\theta \in \Theta : \mathbb{W}^n_\theta(\mathscr{B}|\mathbf{t}, \hat{\theta}) > 1 - \epsilon\}|\hat{\boldsymbol{\theta}} = \hat{\theta}\right) \geq 1 - \gamma$.

June 9, 2018  To appear in IEEE Transactions on Information Theory

*Proposition 2.3:* A set $\Lambda \subset \Theta$ is called a *confidence set* for $\Theta$ if $\Pr(\boldsymbol{\theta} \notin \Lambda | \hat{\boldsymbol{\theta}} = \hat{\theta}) < \gamma$, where $\gamma$ denotes the confidence level. If $\Lambda$ is a confidence set of level $\gamma$ and $\mathscr{B}$ is a common $\eta$-image of the collection of DMCs $\{\mathbb{W}_\theta : \mathscr{T} \times \Theta \longmapsto \mathscr{Y} \times \mathscr{V}\}_{\theta \in \Lambda}$, then $\mathscr{B}$ is also a robust $\epsilon$-decoding set with $\epsilon = 1 - \eta$.

The statement follows from the fact that any conditional PD is $\Theta$-measurable and from basic properties of measurable functions (see [55, p. 185]).

*Definition 2.4 (Robust I-typical sets):* Robust $\epsilon$-decoding sets can be implemented by introducing the concept of *robust I-typical sets*. A robust I-typical set is defined as

$$\mathscr{T}^n_{[Y_\Lambda V_\Lambda | T\hat{\theta}]_{\delta_n}}(\mathbf{t}) \doteq \bigcup_{\theta \in \Lambda} \mathcal{T}^n_{[Y_\theta V_\theta | T\hat{\theta}]_{\delta_n}}(\mathbf{t}, \hat{\theta}),$$

for an arbitrary subset $\Lambda \subset \Theta$ and $\delta$-sequence $(\delta_n)_{n \in \mathbb{N}_+}$.

*Lemma 2.5:* Given $0 < \gamma, \epsilon < 1$, a necessary and sufficient condition for a robust I-typical set $\mathscr{T}^n_{[Y_\Lambda V_\Lambda | T\hat{\theta}]_{\delta_n}}$ to be a robust $\epsilon$-decoding set is that $\Lambda$ be a confidence set of level $\gamma$.

*Proof:* We start proving the necessary part of this condition, namely $\Pr\left(\Lambda | \hat{\boldsymbol{\theta}} = \hat{\theta}\right) \geq 1 - \gamma$ (i.e. $\Lambda$ is a confidence set) implies $\Pr\left(\Lambda_\epsilon | \hat{\boldsymbol{\theta}} = \hat{\theta}\right) \geq 1 - \gamma$ with $\Lambda_\epsilon \doteq \{\theta \in \Theta : \mathbb{W}^n_\theta\left(\mathscr{T}^n_{[Y_\Lambda V_\Lambda | T\hat{\theta}]_{\delta_n}} | \mathbf{t}, \hat{\theta}\right) > 1 - \epsilon\}$ (i.e. $\mathscr{T}^n_{[Y_\Lambda V_\Lambda | T\hat{\theta}]_{\delta_n}}$ is a robust $\epsilon$-decoding set). From Proposition 1.4-(ii) it is easy to see that $\mathscr{T}^n_{[Y_\Lambda V_\Lambda | T\hat{\theta}]_{\delta_n}}$ is a common $\eta$-image for the collection of DMCs $\mathcal{W}_\Lambda$ (with $\eta = 1 - \epsilon$), and thus the proof follows as a consequence of Proposition 2.3. In order to prove the sufficiency condition, we need show that if $\mathscr{T}^n_{[Y_\Lambda V_\Lambda | T\hat{\theta}]_{\delta_n}}$ is a robust $\epsilon$-decoding set then $\Lambda$ must be a confidence set of level $\gamma$. Instead of this, we shall show the converse implication, namely $\Pr\left(\Lambda | \hat{\boldsymbol{\theta}} = \hat{\theta}\right) < 1 - \gamma$ (i.e. $\Lambda$ is not a confidence set) implies that $\Pr\left([\Lambda_\epsilon]^c | \hat{\boldsymbol{\theta}} = \hat{\theta}\right) \geq 1 - \gamma$ where $[\Lambda_\epsilon]^c = \{\theta \in \Theta : \mathbb{W}^n_\theta\left(\mathscr{T}^n_{[Y_\Lambda V_\Lambda | T\hat{\theta}]_{\delta_n}} | \mathbf{t}, \hat{\theta}\right) \leq \epsilon\}$ (i.e. $\mathscr{T}^n_{[Y_\Lambda V_\Lambda | T\hat{\theta}]_{\delta_n}}$ is not a robust $\epsilon$-decoding set). Actually, from Lemma 2.1 we note that for all $\theta \in [\Lambda]^c \cap \Theta$ there exists $n_0 \in \mathbb{N}_+$ such that $\mathbb{W}^n_\theta\left(\mathscr{T}^n_{[Y_\Lambda V_\Lambda | T\hat{\theta}]_{\delta_n}} | \mathbf{t}, \hat{\theta}\right) \leq \epsilon$ provided by $n \geq n_0$. Consequently, the proof follows immediately by

noting that $[\Lambda]^c \cap \Theta \subseteq [\Lambda_\epsilon]^c$ and $\Pr\left([\Lambda]^c \cap \Theta | \hat{\boldsymbol{\theta}} = \hat{\theta}\right) \geq 1 - \gamma$. ∎

*Theorem 2.6 (Cardinality of robust I-typical sets):* Consider an arbitrary collection of conditional PDs (or channels) $\{\mathbb{W}_\theta : \mathscr{T} \times \Theta \longmapsto \mathscr{Y} \times \mathscr{V}\}_{\theta \in \Lambda}$ together with its associated robust I-typical set $\mathscr{T}^n_{[Y_\Lambda V_\Lambda | T\hat{\theta}]_{\delta_n}}(\mathbf{t}) \doteq \bigcup_{\theta \in \Lambda} \mathfrak{T}^n_{[Y_\theta V_\theta | T\hat{\theta}]_{\delta_n}}(\mathbf{t}, \hat{\theta})$, for all $\mathbf{t} \in \mathscr{T}^n_{[T|\hat{\theta}]_{\delta_n}}$. Then, there exists an index $n_0 \in \mathbb{N}_+$ such that for all $n \geq n_0$ the size of $\mathscr{T}^n_{[Y_\Lambda V_\Lambda | T\hat{\theta}]_{\delta_n}}$ can be bounded as follows:

$$\left| \frac{1}{n} \log \|\mathscr{T}^n_{[Y_\Lambda V_\Lambda | T\hat{\theta}]_{\delta_n}}(\mathbf{t})\| - H(Y_\theta, V_\theta | T, \hat{\boldsymbol{\theta}} = \hat{\theta}) \right| \leq \eta_n,$$

where $H(Y_\Lambda, V_\Lambda | T, \hat{\boldsymbol{\theta}} = \hat{\theta}) \doteq \sup_{\theta \in \Lambda} H(Y_\theta, V_\theta | T, \hat{\boldsymbol{\theta}} = \hat{\theta})$ and $\eta_n \to 0$ as $\delta_n \to 0$ and $n \to \infty$.

The quantity $H(Y_\Lambda, V_\Lambda | T, \hat{\boldsymbol{\theta}} = \hat{\theta})$ may be interpreted as the conditional entropy of the set $\mathcal{W}_\Lambda$ and can be shown to equal the I-projection (cf. [53]) of the uniform PD on the set $\mathcal{W}_\Lambda$. Before proving this theorem we need the following result.

*Lemma 2.7:* Consider an arbitrary set of DMCs $\mathcal{W}_\Lambda \doteq \{\mathbb{W}_\theta : \mathscr{T} \times \Theta \longmapsto \mathscr{Y} \times \mathscr{V}\}_{\theta \in \Lambda}$ together with its associated set of I-typical sequences $\mathscr{T}^n_{[Y_\Lambda V_\Lambda | T\hat{\theta}]_{\delta_n}}(\mathbf{t}) \doteq \bigcup_{\theta \in \Lambda} \mathfrak{T}^n_{[Y_\theta V_\theta | T\hat{\theta}]_{\delta_n}}(\mathbf{t}, \hat{\theta})$ and let the set $\mathscr{T}^n_{[\Sigma]}(\mathbf{t}) \doteq \bigcup_{V_n \in \Sigma} \mathfrak{T}^n_{[V_n]}(\mathbf{t})$ with $\Sigma \doteq \mathcal{W}_\Lambda \cap \mathscr{P}_n(\mathscr{Y} \times \mathscr{V})$, for every $\mathbf{t} \in \mathscr{T}^n$. Then, the size of $\mathscr{T}^n_{[Y_\Lambda V_\Lambda | T\hat{\theta}]_{\delta_n}}$ is bounded as

$$\left| \frac{1}{n} \log \|\mathscr{T}^n_{[Y_\Lambda V_\Lambda | T\hat{\theta}]_{\delta_n}}\| - \max_{V_n \in \Sigma} H(V_n | \hat{P}_n) \right| \leq \|\mathscr{T}\| \|\mathscr{Y}\| \|\mathscr{V}\| n^{-1} \log(1 + n).$$

Furthermore, if the set $\mathcal{W}_\Lambda$ is convex then the upper bound can be replaced by $\|\mathscr{T}^n_{[Y_\Lambda V_\Lambda | T\hat{\theta}]_{\delta_n}}\| \leq \exp\left\{ n \max_{V_n \in \Sigma} H(V_n | \hat{P}_n) \right\}$.

The lower and upper bound for a non convex set $\mathcal{W}_\Lambda$ can be easily proved, while for the a convex set this follows straightforward as a generalization from results found in [57].

*Proof: Theorem 2.6.* We first show that the size of $\mathscr{T}^n_{[Y_\Lambda V_\Lambda | T\hat{\theta}]_{\delta_n}}$ is asymptotically equal to the size of $\mathscr{T}^n_{[\Sigma]}(\mathbf{t}) = \bigcup_{V_n \in \Sigma} \mathfrak{T}^n_{[V_n]}(\mathbf{t})$, where $\Sigma = \mathcal{W}_\Lambda \cap \mathscr{P}_n(\mathscr{Y} \times \mathscr{V})$ is the intersection of $\mathcal{W}_\Lambda$ with the set $\mathscr{P}_n(\mathscr{Y} \times \mathscr{V})$ of empirical distributions induced by sequences of length $n$. In particular, there exists an index $n_0$ such that for all $n \geq n_0$ and $\mathbf{t} \in \mathcal{T}^n_{[T|\hat{\theta}]_{\delta_n}}$,

$$\|\mathscr{T}^n_{[\Sigma]}(\mathbf{t})\| \leq \|\mathscr{T}^n_{[Y_\Lambda V_\Lambda|T\hat{\theta}]_{\delta_n}}(\mathbf{t})\| \leq (1+n)^{\|\mathscr{T}\|\|\mathscr{Y}\|\|\mathscr{V}\|}\|\mathscr{T}^n_{[\Sigma]}(\mathbf{t})\|. \quad (53)$$

The lower bound in (53) is obvious. Let us assume that there exists a sequence of $(\epsilon_n)_{n \in \mathbb{N}_+}$ such that for all $n \geq n_0$ and each $\mathbf{t} \in \mathscr{T}^n$,

$$\bigcup_{\theta \in \Lambda} \mathcal{T}^n_{[Y_\theta V_\theta|T\hat{\theta}]_{\delta_n}}(\mathbf{t}) \subseteq \bigcup_{V_n \in \Sigma} \mathscr{T}^n_{[V_n]_{\epsilon_n}}(\mathbf{t}), \quad (54)$$

from which the upper bound in (53) follows as

$$\left\|\bigcup_{\theta \in \Lambda} \mathscr{T}^n_{[Y_\theta V_\theta|T\hat{\theta}]_{\delta_n}}(\mathbf{t})\right\| \overset{(a)}{\leq} \sum_{V_n \in \Sigma} \|\mathscr{T}^n_{[V_n]_{\epsilon_n}}(\mathbf{t})\|$$

$$\overset{(b)}{\leq} (1+n)^{\|\mathscr{T}\|\|\mathscr{Y}\|\|\mathscr{V}\|}\|\mathscr{T}^n_{[\Sigma]}(\mathbf{t})\|, \quad (55)$$

where (a) follows from (54) and the union bound, (b) follows from $\|\mathscr{T}^n_{[V_n]_{\epsilon_n}}(\mathbf{t})\| \leq (1+n)^{\|\mathscr{T}\|\|\mathscr{Y}\|\|\mathscr{V}\|}\|\mathscr{T}^n_{[V_n]}(\mathbf{t})\|$ and the fact that for every $V_n(\cdot|t), \bar{V}_n(\cdot|t) \in \Sigma$ with $V_n(\cdot|t) \neq \bar{V}_n(\cdot|t)$ and each $\mathbf{t} \in \mathscr{T}^n$ we have $\mathcal{T}^n_{[V_n]}(\mathbf{t}) \cap \mathcal{T}^n_{[\bar{V}_n]}(\mathbf{t}) = \{\emptyset\}$.

Now we turn to prove statement (54). First of all, note that $\mathcal{W}_\Lambda$ is a relatively $\tau_0$-open subset of $\mathcal{W}_\Lambda \cup \mathscr{P}_n(\mathscr{Y} \times \mathscr{V})$ in the $\tau_0$-topology[3] [36], i.e., every $W \in \mathcal{W}_\Lambda$ has a $\tau_0$-neighborhood such that the $\varepsilon$-open ball $U_0(W, \mathscr{P}, \varepsilon) \subset \mathcal{W}_\Lambda$. Hence, given an arbitrary sequence $(\epsilon_n)_{n \in \mathbb{N}_+}$ the $\varepsilon_n$-open ball satisfies $U_0(W, \mathscr{P}, \varepsilon_n) \cap \mathscr{P}_n(\mathscr{Y} \times \mathscr{V}) \subset \mathcal{W}_\Lambda$ with large enough $n$. As a consequence, there exists an index $n_0 \in \mathbb{N}_+$ such that for all $n \geq n_0$, we can choose the sequence $\varepsilon'_n \doteq (\|\mathscr{Y}\|\|\mathscr{V}\|)^{-1}[\epsilon_n + \sqrt{\delta_n/2}\log\left(\sqrt{\delta_n/2}/(\|\mathscr{T}\|\|\mathscr{Y}\|^2\|\mathscr{V}\|^2)\right)]$ and pick a conditional PD $V_n(\cdot|t) \in U_0(W, \mathscr{P}, \varepsilon_n) \cap \mathscr{P}_n(\mathscr{Y} \times \mathscr{V})$ with $V_n(\cdot|t) \gg W(\cdot|t)$ such that $W(y, v|t)\log\dfrac{W(y, v|t)}{V_n(y, v|t)} \leq \varepsilon'_n$ for all $(y, v) \in \mathscr{Y} \times \mathscr{V}$ and $\hat{P}_n(t) > 0$. On the other hand, observe that any sequence $(\mathbf{y}, \mathbf{v}) \in$ $\mathfrak{T}^n_{[YV|T\hat{\theta}]_{\delta_n}}(\mathbf{t})$ implies $\mathcal{D}(\widehat{W}_n\|W|\hat{P}_n) \leq \delta_n$ with $\widehat{W}_n \ll W$ and by continuity Lemma 1.5 this leads to $\mathcal{D}(\widehat{W}_n\|V_n|\hat{P}_n) \leq \mathcal{D}(W\|V_n|\hat{P}_n) - \sqrt{\delta_n/2}\log\left(\sqrt{\delta_n/2}/(\|\mathscr{T}\|\|\mathscr{Y}\|^2\|\mathscr{V}\|^2)\right)$. Then, it is easy to see that $\mathcal{D}(\widehat{W}_n\|V_n|\hat{P}_n) \leq \epsilon_n$ and therefore $(\mathbf{y},\mathbf{v}) \in \mathfrak{T}^n_{[V_n]_{\epsilon_n}}(\mathbf{t})$. This proves that for each $W \in \mathcal{W}_\Lambda$ and sufficiently large $n$, it is possible to find a conditional PD $V_n(\cdot|t) \in \Sigma$ and a sequence $(\epsilon_n)_{n\in\mathbb{N}_+}$ such that $\mathfrak{T}^n_{[YV|T\hat{\theta}]_{\delta_n}}(\mathbf{t}) \subseteq \mathfrak{T}^n_{[V_n]_{\epsilon_n}}(\mathbf{t})$, which establishes (54).

---

[3]The $\tau_0$-topology on $\mathscr{P}(\mathscr{Y} \times \mathscr{V})$ si defined by the basic neighborhoods $U_0(W, \mathscr{P}, \varepsilon) = \{V(\cdot|t) \in \mathscr{P}(\mathscr{Y} \times \mathscr{V}) : |W(y, v|t) - V(y, v|t)| < \epsilon, \ V(y, v|t) = 0 \text{ if } W(y, v|t) = 0, \text{ for all } P_T(t) > 0\}$.

Using similar arguments as above and the uniform continuity of the entropy function, it can be shown that there exists $n'_0 \in \mathbb{N}_+$ and $(\xi'_n)_{n\in\mathbb{N}_+}$ such that for all $n \geq n'_0$ and each $\mathbf{t} \in \mathfrak{T}^n_{[T|\hat{\theta}]_{\delta_n}}$,

$$\left|\max_{V_n \in \Sigma} H(V_n|\hat{P}_n) - \sup_{\theta \in \Lambda} H(Y_\theta, V_\theta|T, \hat{\boldsymbol{\theta}} = \hat{\theta})\right| \leq \xi'_n, \tag{56}$$

with $\xi'_n \to 0$ as $n \to \infty$. Finally, the theorem follows by combining inequalities (53) with Lemma 2.7 and inequalities (56), and by setting $\eta_n \doteq \xi'_n + 2\|\mathscr{T}\|\|\mathscr{Y}\|\|\mathscr{V}\|n^{-1}\log(n+1)$, for all $n \geq \max\{n'_0, n_0\}$. ∎

## APPENDIX III

### INFORMATION INEQUALITIES

Given arbitrary measurable functions $\{f_k : \mathscr{Y} \times \mathscr{V} \mapsto \mathbb{C}\}_{k=1}^K$ and numbers $\{\lambda_k \in \mathbb{C}\}_{k=1}^K$, the set $\mathcal{L} = \{\mathbb{W}_\theta(y,v|t,\hat{\theta}) \in \mathscr{P}(\mathscr{Y} \times \mathscr{V}) : \sum_{y\in\mathscr{Y}}\sum_{v\in\mathscr{V}} f_k(y,v)\mathbb{W}_\theta(y,v|t,\hat{\theta}) = \lambda_k, \ 1 \leq k \leq K \text{ and } t \in \mathscr{T}\}$ if non-empty, is called a *linear family* of conditional PDs [40].

*Theorem 3.1:* For an arbitrary set of states $\Lambda \subset \Theta$, let $\mathcal{W}_\Lambda \doteq \{\mathbb{W}_\theta : \mathscr{T} \times \Theta \longmapsto \mathscr{Y} \times \mathscr{V}\}_{\theta\in\Lambda} \subset \mathscr{P}(\mathscr{Y}\times\mathscr{V})$ be a convex set of conditional PDs (or channels) with finite input, state and output alphabets $(\mathscr{T},\mathscr{V},\mathscr{Y})$ and let $\mathbb{W}_{\theta^\star}(y,v|t,\hat{\theta}) \in \mathcal{W}_\Lambda$ be the channel such that $S_P(\mathbb{W}_{\theta^\star}) = S_P(\mathcal{W}_\Lambda)$, respect to a PD $q_{TX|U\hat{\theta}} \in \mathscr{P}_\Gamma$ as defined in (10). Then the following inequality holds

$$\inf_{\lambda \in \Lambda} I(T;Y_\lambda|V_\lambda,\hat{\theta}) \leq I(T;Y_\theta|V_\theta) - \left[\mathcal{D}\big((Y_\theta,V_\theta)\|(Y_{\theta^\star},V_{\theta^\star})|T,\hat{\theta}\big) - \mathcal{D}\big((Y_\theta,V_\theta)\|(Y_{\theta^\star},V_{\theta^\star})|\hat{\theta}\big)\right], \tag{57}$$

for every state $\theta \in \Lambda$. Furthermore, if the asserted inequality holds for some $\theta^\star \in \Lambda$ and all $\theta \in \Lambda$, then $\theta^\star$ must provide the infimum value of the mutual information over the set $\Lambda$, i.e., $I_\epsilon(T; Y_{\theta^\star} | V_{\theta^\star}, \hat{\theta}) = \inf_{\theta \in \Lambda} I(T; Y_\theta | V_\theta, \hat{\theta}) + \epsilon$ with $\epsilon > 0$. Moreover, the inequality (57) is actually an equality if $\mathcal{W}_\Lambda$ is a *linear family* of conditional PDs ($\mathcal{W}_\Lambda \subset \mathcal{L}$).

*Proof:* Let $\epsilon > 0$ and $\mathbb{W}_{\theta^\star}(y, v | t, \hat{\theta})$ with $\theta^\star \in \Lambda$ be the channel state that yields to $I_\epsilon(T; Y_{\theta^\star} | V_{\theta^\star}, \hat{\theta}) = \inf_{\lambda \in \lambda} I(T; Y_\lambda | V_\lambda, \hat{\theta}) + \epsilon$. For arbitrary $\mathbb{W}_\theta(y, v | t, \hat{\theta})$ with state $\theta \in \Lambda$, the convexity of $\mathcal{W}_\Lambda$ guarantees that $\mathbb{W}^{(\alpha)}_{\theta, \theta^\star}(y, v | t, \hat{\theta}) = (1 - \alpha) \mathbb{W}_{\theta^\star}(y, v | t, \hat{\theta}) + \alpha \mathbb{W}_\theta(y, v | t, \hat{\theta}) \in \mathcal{W}_\Lambda$ for all $\alpha \in [0, 1]$. Observe that $\mathbb{W}^{(\alpha)}_{\theta, \theta^\star}(y, v | t, \hat{\theta})$ is linear in $\alpha$ and $I(T; Y_\theta | V_\theta, \hat{\theta})$ is a convex function in $\mathbb{W}_\theta(y, v | t, \hat{\theta})$, which implies that $I(T; Y^{(\alpha)}_{\theta, \theta^\star} | V^{(\alpha)}_{\theta, \theta^\star}, \hat{\theta})$ is a convex function in $\alpha$. Hence, the difference quotient of $I(T; Y^{(\alpha)}_{\theta, \theta^\star} | V^{(\alpha)}_{\theta, \theta^\star}, \hat{\theta})$ evaluated in $\alpha = 0$ is given by,

$$\Delta_t(\alpha = 0) = \frac{1}{t} \big[ I(T; Y^{(t)}_{\theta, \theta^\star} | V^{(t)}_{\theta, \theta^\star}, \hat{\theta}) - I_\epsilon(T; Y_{\theta^\star} | V_{\theta^\star}, \hat{\theta}) \big], \tag{58}$$

with $\Delta_t(\alpha = 0) \geq 0$ for each $t \in (0, 1]$. Thus, there exits some $\tilde{t} \in (0, t)$ such that

$$0 \leq \Delta_t(\alpha = 0) = \frac{\partial}{\partial \alpha} I(T; Y^{(\alpha)}_{\theta, \theta^\star} | V^{(\alpha)}_{\theta, \theta^\star}, \hat{\theta}) \Big|_{\alpha = \tilde{t}}. \tag{59}$$

While,

$$\frac{\partial}{\partial \alpha} I(T; Y^{(\alpha)}_{\theta, \theta^\star} | V^{(\alpha)}_{\theta, \theta^\star}, \hat{\theta}) = \sum_{t \in \mathcal{T}} \sum_{v \in \mathcal{V}} \sum_{y \in \mathcal{Y}} q_{T|\hat{\theta}}(t | \hat{\theta}) \big[ \mathbb{W}_\theta(y, v | t, \hat{\theta}) - \mathbb{W}_{\theta^\star}(y, v | t, \hat{\theta}) \big] \log \frac{\mathbb{W}^{(\alpha)}_{\theta, \theta^\star}(y, v | t, \hat{\theta})}{\mathbb{W}^{(\alpha)}_{\theta, \theta^\star} q_{T|\hat{\theta}}(y, v | \hat{\theta})}, \tag{60}$$

and by taking $t \to 0$ in expression (59), we obtain

$$\begin{aligned} 0 &\leq \lim_{\tilde{t} \to 0} \Delta_t(\alpha = 0) = \lim_{\tilde{t} \to 0} \frac{\partial}{\partial \alpha} I(T; Y^{(\alpha)}_{\theta, \theta^\star} | V^{(\alpha)}_{\theta, \theta^\star}, \hat{\theta}) \Big|_{\alpha = \tilde{t}}, \\ &= \sum_{t \in \mathcal{T}} \sum_{v \in \mathcal{V}} \sum_{y \in \mathcal{Y}} P_{T|\hat{\theta}}(t | \hat{\theta}) \big[ \mathbb{W}_\theta(y, v | t, \hat{\theta}) - \mathbb{W}_{\theta^\star}(y, v | t, \hat{\theta}) \big] \log \frac{\mathbb{W}_{\theta^\star}(y, v | t, \hat{\theta})}{\mathbb{W}_{\theta^\star} q_{T|\hat{\theta}}(y, v | \hat{\theta})}, \\ &= I(T; Y_\theta | V_\theta) + \mathcal{D}\big((Y_\theta, V_\theta) \| (Y_{\theta^\star}, V_{\theta^\star}) | \hat{\theta}\big) - \mathcal{D}\big((Y_\theta, V_\theta) \| (Y_{\theta^\star}, V_{\theta^\star}) | T, \hat{\theta}\big) - I_\epsilon(T; Y_{\theta^\star} | V_{\theta^\star}, \hat{\theta}), \end{aligned} \tag{61}$$

where we have used the fact that $S_p(\mathbb{W}_\theta) \subseteq S_p(\mathbb{W}_{\theta^\star})$. Since expression (61) is always positive this concludes the proof of the inequality (57). In order to show the equality, observe that under the assumption that $\mathcal{W}_\Lambda$ is a linear family of conditional PDs. For every $\mathbb{W}_\theta(y,v|t,\hat{\theta}) \in \mathcal{L}$, there is some $\alpha < 0$ such that $\mathbb{W}_{\theta,\theta^\star}^{(\alpha)}(y,v|t,\hat{\theta}) = (1-\alpha)\mathbb{W}_{\theta^\star}(y,v|t,\hat{\theta}) + \alpha \mathbb{W}_\theta(y,v|t,\hat{\theta}) \in \mathcal{L}$. Therefore, we must have $(\partial/\partial\alpha) I(T; Y_{\theta,\theta^\star}^{(\alpha)}|V_{\theta,\theta^\star}^{(\alpha)}, \hat{\theta})\big|_{\alpha=0} = 0$, i.e.

$$\sum_{t \in \mathcal{T}} \sum_{v \in \mathcal{V}} \sum_{y \in \mathcal{Y}} q_{T|\hat{\theta}}(t|\hat{\theta}) \big[\mathbb{W}_\theta(y,v|t,\hat{\theta}) - \mathbb{W}_{\theta^\star}(y,v|t,\hat{\theta})\big] \log \frac{\mathbb{W}_{\theta^\star}(y,v|t,\hat{\theta})}{\mathbb{W}_{\theta^\star} q_{T|\hat{\theta}}(y,v|\hat{\theta})} = 0,$$

for all $\mathbb{W}_\theta(y,v|t,\hat{\theta}) \in \mathcal{L}$, and this proves the equality in (57). ∎

## APPENDIX IV

### EVALUATION OF SOME INDEFINITE INTEGRALS

In this Appendix we want to evaluate the following indefinite integral defined by (45)

$$I(r_{\text{opt}}, \phi_\mathcal{E}) = \int_{r_{\text{opt}}}^{\infty} \int_{\phi_{\hat{H}} - \phi_\mathcal{E}}^{\phi_{\hat{H}} + \phi_\mathcal{E}} \mathbb{D}\, r \exp\left(-\mathbb{A} r^2 + \mathbb{B} r \cos(\phi_H - \phi_{\mu_{\hat{H}}})\right) dr d\phi_H, \qquad (62)$$

where $r_{\text{opt}}, \mathbb{A}, \mathbb{B} \in \mathbb{R}_+$ with $\mathbb{D} \doteq \frac{\mathbb{A}}{\pi} \exp\left(-\frac{\mathbb{B}^2}{4\mathbb{A}}\right)$ and $\phi_\mathcal{E} \in [-\pi, \pi]$. First, we use integration by parts and the series expansion of [58, Eq. 6.9] to obtain

$$\int_{\phi_{\hat{H}} - \phi_\mathcal{E}}^{\phi_{\hat{H}} + \phi_\mathcal{E}} \exp\left(\mathbb{B} r \cos(\phi_H - \phi_{\mu_{\hat{H}}})\right) d\phi_H = 2 \int_0^{\phi_\mathcal{E}} \left[I_0(r) + 2 \sum_{k=1}^{\infty} I_k(r) \cos(k\phi_H)\right] d\phi_H,$$

$$= 2 I_0(\mathbb{B} r) \phi_\mathcal{E} + 4 \sum_{n=1}^{\infty} I_n(\mathbb{B} r) \left(\frac{\sin(k\phi_\mathcal{E})}{k}\right), \quad (63)$$

where $I_k(x)$ is the k-th order *modified Bessel function of the first kind* [42, Eq. (8.445)]

$$I_n(x) \doteq \sum_{k=1}^{\infty} \frac{(-1)^k}{k!\, \Gamma(n+k+1)} \left(\frac{x}{2}\right)^{n+2k}, \qquad (64)$$

and $\Gamma(\cdot)$ is the Gamma function. We now compute the remainder term given by

$$\int_{r_{\text{opt}}}^{\infty} r \exp\left(-\mathbb{A} r^2\right) I_n(\mathbb{B} r) dr = \frac{1}{2\mathbb{A}} \exp\left(\frac{\mathbb{B}^2}{4\mathbb{A}}\right) \mathcal{Q}_{1,k}\left(\frac{\mathbb{B}}{\sqrt{2\mathbb{A}}}, \sqrt{2\mathbb{A}} r_{\text{opt}}\right), \qquad (65)$$

and $\mathcal{Q}_{1,n}(\alpha, \beta)$ is the *Nutall Q-function* defined by [59]

$$\mathcal{Q}_{1,n}(\alpha, \beta) \doteq \int_\beta^\infty x \exp\left(-\frac{x^2 + \alpha^2}{2}\right) I_n(\alpha\, x) dx, \tag{66}$$

with non-negative reals $\alpha, \beta$ (see [60] for its numerical evaluation). Actually, the integral in (62) follows from (63) and (65),

$$\begin{aligned}
I(r_{\text{opt}}, \phi_\mathcal{E}) &= \frac{1}{\pi} \Bigg[ \mathcal{Q}_1\left(\frac{\mathbb{B}}{\sqrt{2\mathbb{A}}}, \sqrt{2\mathbb{A}} r_{\text{opt}}\right) \phi_\mathcal{E} \\
&\quad + 2 \sum_{k=1}^\infty \mathcal{Q}_{1,k}\left(\frac{\mathbb{B}}{\sqrt{2\mathbb{A}}}, \sqrt{2\mathbb{A}} r_{\text{opt}}\right) \left(\frac{\sin(k\phi_\mathcal{E})}{k}\right) \Bigg],
\end{aligned} \tag{67}$$

where $\mathcal{Q}_1(\alpha, \beta) = \mathcal{Q}_{1,1}(\alpha, \beta)$ is the first-order *Marcum Q-function* [43]. This infinite sum does not seem to be amenable to further simplifications yielding a closed-form expression. Numerical simulations showed that it can be well-approximated using only two terms, i.e.,

$$\begin{aligned}
I(r_{\text{opt}}, \phi_\mathcal{E}) &\approx \frac{1}{\pi} \Bigg[ \mathcal{Q}_1\left(\frac{\mathbb{B}}{\sqrt{2\mathbb{A}}}, \sqrt{2\mathbb{A}} r_{\text{opt}}\right) \phi_\mathcal{E} \\
&\quad + 2\mathcal{Q}_{1,1}\left(\frac{\mathbb{B}}{\sqrt{2\mathbb{A}}}, \sqrt{2\mathbb{A}} r_{\text{opt}}\right) \sin(\phi_\mathcal{E}) \Bigg].
\end{aligned} \tag{68}$$

The evaluation of the expectation in (41) is obtained by computing the following integral

$$\begin{aligned}
I(A, B, P) &= \frac{1}{P} \int_0^\infty x \log_2(A + Bx) \exp\left(-\frac{x^2}{2P}\right) dx, \\
&= \log_2(A) + \exp\left(\frac{B}{AP}\right) E_1\left(\frac{B}{AP}\right),
\end{aligned} \tag{69}$$

where $E_1(z) = \int_z^\infty t^{-1} \exp(-t) dt$ denotes the *exponential integral function*.

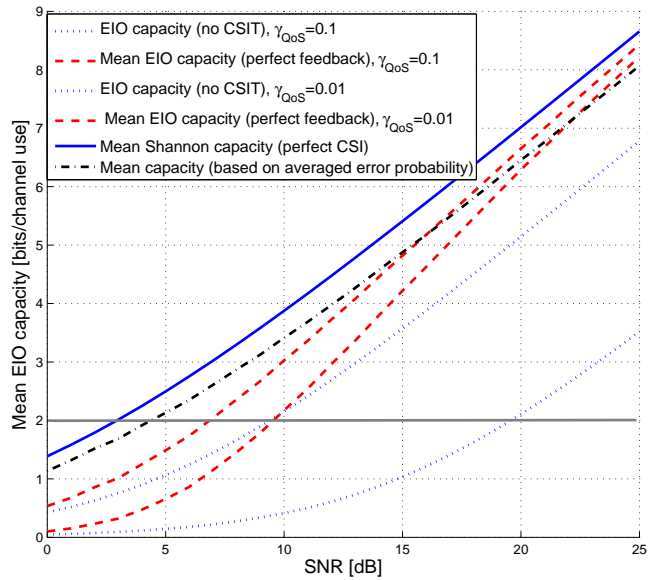

(a) Mean EIO capacity with perfect feedback (dashed lines), without feedback (dotted lines), mean capacity based on the averaged error probability (dashed-dot line) and mean Shannon capacity (solid line) vs. SNR, for $\gamma \in \{0.1, 0.01\}$ and $N = 1$.

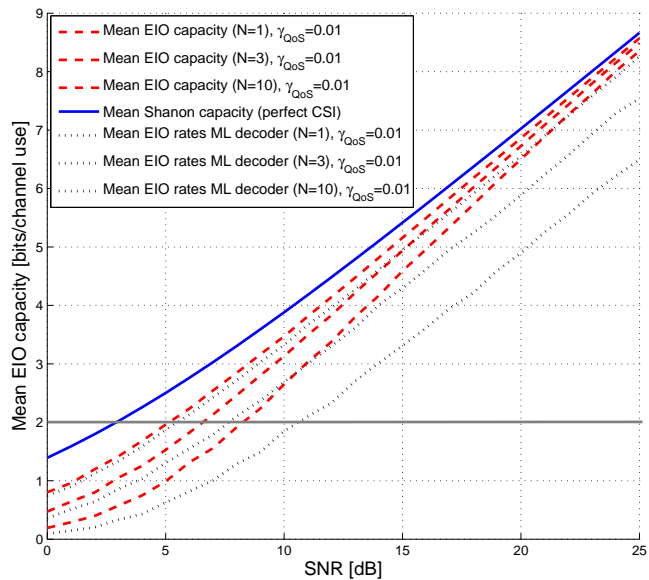

(b) Mean EIO capacity with perfect feedback (dashed lines), achievable EIO rates associated to the mismatched ML decoder (dotted lines) and mean Shannon capacity (solid line) vs. SNR, for $\gamma = 0.01$ and $N \in \{1, 3, 10\}$.

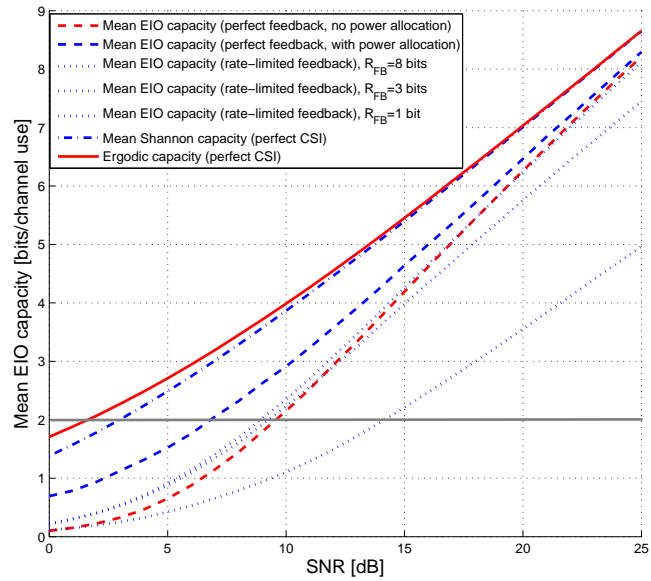

(c) Mean EIO capacity with perfect feedback (with and without power allocation, dashed line), with rate-limited feedback $R_{\text{FB}} \in \{1, 3, 8\}$ (power allocation, dotted lines), ergodic capacity (solid line) and mean Shannon capacity (dashed-dot line) vs. SNR, for $\gamma = 0.01$ and $N = 1$.

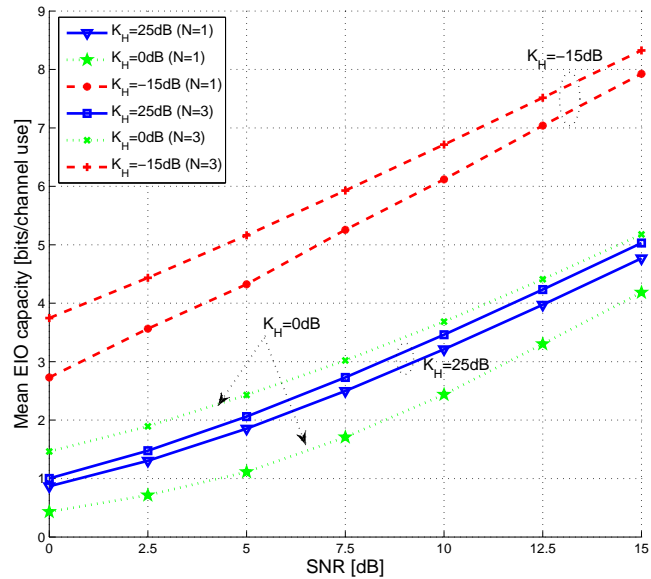

(d) Mean EIO capacity for different Rice factors $K_H \in \{-15, 0, 25\}$dB and amounts of training $N \in \{1, 3\}$ with perfect feedback vs. SNR.